\documentclass[twocolumn,useAMS,usenatbib]{mn2e}
\usepackage{graphicx}
\usepackage{natbib}
\usepackage{bm}
\usepackage{amsfonts}
\usepackage{latexsym}
\usepackage[latin1]{inputenc}
\usepackage{amsmath}
\usepackage{epsfig}
\usepackage{amsbsy}
\usepackage{psfrag}
\usepackage{mnfixes}

\newcommand{\mtb}[1]{\mathbf{#1}}

\newcommand{\mtv}[1]{\textbf{\emph{#1}}}

\newcommand{\veps}{\varepsilon}

\def \nn  {\nonumber}

\def \eps{\epsilon}
\def \veps{\varepsilon}
\def\jnl@style{\rm}
\def\aaref@jnl#1{{\jnl@style#1}}

\def\aaref@jnl#1{{\jnl@style#1}}

\def\aj{\aaref@jnl{AJ}}                   
\def\apj{\aaref@jnl{ApJ}}                 
\def\apjl{\aaref@jnl{ApJ}}                
\def\apjs{\aaref@jnl{ApJS}}               
\def\apss{\aaref@jnl{Ap\&SS}}             
\def\aap{\aaref@jnl{A\&A}}                
\def\aapr{\aaref@jnl{A\&A~Rev.}}          
\def\aaps{\aaref@jnl{A\&AS}}              
\def\mnras{\aaref@jnl{MNRAS}}             
\def\prd{\aaref@jnl{Phys.~Rev.~D}}        
\def\prl{\aaref@jnl{Phys.~Rev.~Lett.}}    
\def\qjras{\aaref@jnl{QJRAS}}             
\def\skytel{\aaref@jnl{S\&T}}             
\def\ssr{\aaref@jnl{Space~Sci.~Rev.}}     
\def\zap{\aaref@jnl{ZAp}}                 
\def\nat{\aaref@jnl{Nature}}              
\def\aplett{\aaref@jnl{Astrophys.~Lett.}} 
\def\apspr{\aaref@jnl{Astrophys.~Space~Phys.~Res.}} 
\def\physrep{\aaref@jnl{Phys.~Rep.}}      
\def\physscr{\aaref@jnl{Phys.~Scr}}       

\title[Low T/W instability and the co-rotation point]{The intimate relation between the low T/W instability and the co-rotation point}

\author[A. Passamonti and N. Andersson]
{A. Passamonti$^1$\thanks{E-mail:andrea.passamonti@oa-roma.inaf.it} , N. Andersson$^2$ \\ \\
$^1$ INAF-Osservatorio Astronomico di Roma, via Frascati 44, I-00040, Monteporzio Catone (Roma), Italy \\
$^2$School of Mathematics and STAG Research Centre, University of Southampton, Southampton SO17 1BJ, UK}

\begin{document}

\date{\today}

\pagerange{\pageref{firstpage}--\pageref{lastpage}} \pubyear{}
\maketitle

\label{firstpage}


\begin{abstract}

We study the low T/W instability associated with the f-mode of differentially rotating stars. 
Our stellar models are described by a polytropic equation of state and the rotation profile is given by the standard j-constant  law. 
 The properties of the relevant oscillation modes, including the instability growth time, are 
determined from time evolutions of the linearised dynamical equations in Newtonian gravity. 
In order to analyse the instability we monitor also the canonical energy and angular momentum. 
Our results demonstrate that the  $l=m=2$ f-mode becomes unstable as soon as a co-rotation point develops inside the star (i.e. whenever there is a point  where the mode's pattern speed 
matches the bulk angular velocity). Considering various degrees of differential rotation, we show that the instability grows faster deep inside the co-rotation region and deduce
an empirical relation that correlates 
the mode frequency and the star's parameters,  which captures the main features of the $l=m=2$ f-mode growth time. 
This function is proportional to the product of the kinetic to gravitational energy ratio and the gradient of 
the star's spin, strengthening further the relationship between the co-rotation point and the low T/W instability. 
We briefly consider also the $l=m=2$ r-mode and demonstrate that 	it never moves far inside the co-rotation region even for 
significant differential rotation.
\end{abstract}











\section{Introduction}

Rotating neutron stars may experience a host of non-axisymmetric  instabilities, acting either on a dynamical or secular (e.g. associated with some dissipation mechanism) timescale. It is important to understand the range of possible instabilities, whether they are likely to act in a realistic astrophysical setting and possible repercussions on the evolution of the star. The most pressing issues have a direct connection with observations. In particular, because the unstable oscillation modes are non-axisymmetric they  generate gravitational radiation and a number of plausible scenarios where this radiation may be detectable by advanced interferometers have been suggested. Presently, the most interesting scenarios involve dynamical instabilities triggered during the late stages of core-collapse or in the immediate aftermath of binary merger and gravitational-wave driven instabilities of  more mature stars (associated with either the star's fundamental f-mode or the inertial r-mode). The main topic of the present work, the so-called low T/W instability, is a relative late-comer to the discussion. Just like its classic cousins the dynamical bar-mode instability and the secular instabilities due to viscosity or gravitational radiation, this instability acts on the f-mode of the star. The main distinction is that the low T/W instability requires the star to be differentially rotating. The aim of this paper is to explore, in more detail than in previous work, the nature of this instability. In particular, we want to test the assertion that the instability becomes active as soon as the f-mode enters the co-rotation region (whenever there exists a point in the star where the mode appears stationary with respect to an inertial observer). 
 
The study of instabilities of rotating stars has a long history. Work on Newtonian incompressible self-graviting rotating ellipsoids, like the Maclaurin spheroids, provided the first instability criteria in terms of the parameter $\beta = T/|W| $, where $T$ and $W$ are, respectively, the  kinetic and gravitational energy of the star (note that we will often omit the absolute values in referring to the low T/W instability in the text). 
Early work established that the dynamical bar-mode instability sets in when a star reaches   $\beta_{d} = 0.27$~\citep{1969efe..book.....C}. 
For higher spins than this, a star can reach a new equilibrium configuration with lower energy, maintaining the same mass, volume and angular momentum. The instability is dynamical because it does not require any additional mechanisms, like internal viscosity or the coupling to radiation, to operate. 
Although the models that were considered early on were simplistic the results have been found to be robust. Recent relativistic numerical simulations of rotating stars with both polytropic and more realistic equation of state (EoS) 
have only adjusted this instability threshold slightly to  $\beta_{d} = 0.24-0.25$~\citep{2000ApJ...542..453S, 2007PhRvD..75d4023B}. However, this level of rotation cannot be achieved by a compressible star in uniform rotation, since the corresponding  mass shedding limit is reached for $\beta \simeq 0.1$. To reach higher values of $\beta$ the system must rotate differentially. However, it is still not clear that astrophysical systems will ever reach such large values of $\beta$, at least not for a sustained period (there is evidence of significant differential rotation being generated during core collapse but this phase is interrupted by the core bounce so any related instability do not have much time to grow).

The introduction of differential rotation brings new features to the problem. In particular, numerical simulations led to the discovery that  
stars with a higher degree of differential rotation can suffer dynamical instabilities 
even at much lower values, $\beta \simeq 0.14 $~\citep{2001ApJ...550L.193C}. 
The existence of this instability, now referred to as the low T/W instability,  has been confirmed by  a number of full
3D  numerical simulations as well as in linear analyses. In some cases, the instability threshold has been shown to be as low as  $\beta \simeq 0.01$~\citep{2002MNRAS.334L..27S, 2003MNRAS.343..619S, 2003ApJ...595..352S}. Despite these interesting results, we do not yet have a detailed understanding of the nature of this instability.  It has been proposed that the instability originates from an energy transfer between the bulk motion and an oscillation mode~\citep{2005ApJ...618L..37W}, in a fashion that resembles the shear instability in  thick accretion disks (the so-called Papaloizou-Pringle instability)~\citep{1984MNRAS.208..721P}.  One would expect this energy exchange to occur at the co-rotation point, where the pattern speed 
of a mode matches the local angular velocity of the star (see discussion later). In analogy with accretion disks it was suggested that the star's vortensity  may indicate the  resonant cavity where this instability can grow~\citep{2006ApJ...651.1068O}. However, for rotating stars this criterion does not seem completely satisfactory (at least this has not been demonstrated yet).  

The low T/W instability has been studied
in Newtonian and relativistic non-linear simulations, with different methods and distinct numerical codes.  
There is clear evidence that it may develop during stellar core-collapse~\citep{2005ApJ...625L.119O,2007PhRvL..98z1101O, 2008A&A...490..231S, 2010ApJS..191..439K} and
it has been seen in numerical evolutions of rapidly rotating  cold neutron stars~\citep{2006MNRAS.368.1429S, 2007CoPhC.177..288C, 2010CQGra..27k4104C}. 
Recently, with the development of magneto-hydrodynamical simulations,  the impact of the magnetic field on the dynamical  instabilities has been explored~\citep{2009ApJ...707.1610C, 2013PhRvD..88j4028F, 2014arXiv1405.2144M}.  The results suggest that 
the bar-mode instability can be suppressed by a very strong magnetic field, $B \gtrsim 10^{16}$G.  
Analytical work,  based on a rotating system with cylindrical geometry,  
has shown that the low T/W instability can  be similarly suppressed by a 
strong toroidal magnetic field  $B > 10^{16}$G~\citep{2011MNRAS.413.2207F}. 
Numerical simulations that consider an initial poloidal seed magnetic field, which differential rotation subsequently winds up, have shown that the instability is not affected for poloidal fields  $B < 4 \times10^{13}$G. More interestingly,  
 for $B \geq 5 \times10^{14}$G  a quadrupolar distortion may be amplified by magnetic effects~\citep{2014arXiv1405.2144M}. Much more effort is required in order to explore the interplay between differential rotation, the star's magnetic field and various instabilities. 

In this work we focus on the pure hydrodynamics problem; we ignore magnetic fields, the star's elastic crust, superfluid core etcetera. 
We study the low T/W instability by means of time-evolutions of  the linearised Newtonian equations. 
We focus on stars with a polytropic EoS and assume that the differential rotation is given by the j-constant rotation law. The latter choice is motivated by simulations of both core-collapse and neutron star mergers that suggest that real systems stay rather close to this simple rotation law. 
Our main aim is to determine the instability growth time for varying degrees of differential rotation and connect the results with  
key features like the presence of a co-rotation point in the star.  
The analysis focuses on the $l=m=2$ f-mode (the bar-mode), which in general exhibits the strongest instability. We trace the instability as the degree of differential rotation is increased for a given star, monitoring key indicators like the canonical energy and angular momentum densities.  
This part of the analysis builds on the work of \citet{2006MNRAS.368.1429S}, who considered the canonical angular momentum associated with the unstable mode for cylindrical stars, and 
showed that even though the canonical angular momentum grows due to the instability it  remains zero at the co-rotation point. 
To speed up the numerical simulations and allow us to investigate a large parameter space  we mainly use the Cowling approximation, i.e.
we neglect the perturbation of the gravitational potential. This affects the mode-frequencies at the 20\% level, which should not have significant influence on the qualitative behaviour that is our main focus. 

\section{The problem} \label{sec:Eqs}

\subsection{The Newtonian equations}

As we are considering the problem of a differentially rotating star in Newtonian gravity, we need to solve 
the Euler equation, the continuity equation for  
mass conservation and the Poisson equation for the gravitational potential: 
\begin{eqnarray}
\left( \frac{\partial}{\partial t} + \mtv{v} \cdot \nabla \right) \mtv{v}   & = & -  \nabla \left(  h +  \Phi  \right)
    \, ,  \label{eq:dvdt} \\
\frac{\partial \rho }{\partial t}  & = & - \nabla \cdot  \left(\rho \mtv{v} \right)    \, ,      \label{eq:drhodt} \\
\nabla^2 \Phi & = & 4 \pi G \, \rho \, , \label{eq:dPhi}
\end{eqnarray}
where $G$ is the gravitational constant. In these equations,
the scalar fields $\rho, h$ and $\Phi$ represent, respectively,  the mass
density, the specific enthalpy and the gravitational potential, while $\mtv{v}$
is the fluid velocity. 

To close the system of equations  we need to provide an equation of state. In order to keep things simple, which makes sense as we are mainly interested in qualitative features, we consider a polytropic model;
\begin{equation}
P = k \rho ^{\gamma } \, , \label{eq:polEoS}
\end{equation}
where $k$ is a constant and the adiabatic index is given by
\begin{equation}
\gamma \equiv \frac{d \log P}{ d \log \rho}  = 1 + {1\over n} \, , \label{eq:Gbdef}
\end{equation}
where $n$ is the polytropic index.

For an axisymmetric equilibrium configuration,  equations~(\ref{eq:dvdt})-(\ref{eq:dPhi}) can be written in  integral form~\citep{1986ApJS...61..479H},
\begin{equation}
  h + \Phi + \Psi  =  C \, , \label{eq:bgmu_x}   \\
\end{equation}
where $C$ is an integration constant, and the gravitational potential is determined 
by the integral equation: 
\begin{equation}
 \Phi \left( \mtb{r} \right)  = - G \int_{0}^{\mtb{r}} 
\frac{\rho\left(\mtb{r'} \right)}{ | \mtb{r} - \mtb{r'} |} d \mtb{r'} \label{eq:Pois} \,  .
\end{equation}
In equation (\ref{eq:bgmu_x}),   the centrifugal potential $\Psi$ is  defined 
by:
\begin{equation}
   \Psi  =  - \int \Omega^{2}  \varpi d \varpi  \,  , \label{eq:Psi} \\
\end{equation}
where $\Omega$  is the star's angular velocity  and the cylindrical radius $\varpi = r  \sin \theta$   
represents the distance of a fluid element from the rotation axis. 

In a barotropic fluid, pressure and enthalpy are related by 
\begin{equation}
h = \int  \frac{dP}{\rho}  \, , \label{eq:polEoSb}
\end{equation}
which for a polytropic EoS  leads to:
\begin{equation}
h = \frac{\gamma}{\gamma-1} \frac{P}{\rho} \, . \label{eq:polEoSc}
\end{equation}

\subsection{Linear perturbations} \label{sec:pert-eqs}

In order to explore the unstable oscillations associated with the low T/W instability we need to solve the equations that govern linear perturbations of our differentially rotating configurations. Quite  generally, non-axisymmetric small-amplitude oscillations of a differentially rotating star can be studied 
 with a set of three scalar and one vector perturbation fields, namely 
 the mass density $\delta \rho$, the enthalpy $\delta h$,
the gravitational potential $\delta \Phi$ and the velocity perturbation $\delta \mtv{v}$ .  
These  variables satisfy a system of linearised  equations which, for an inertial observer and in spherical coordinates $[r,\theta,\phi]$, reads
\begin{eqnarray}
\left( \frac{\partial}{\partial t} + \Omega \frac{\partial}{\partial \phi} \right) \delta \mtv{v} & = & 
                     - \nabla \left( \delta h + \delta \Phi \right)  - 2 \mtb{\Omega} \times \delta \mtv{v}
                      \nn \\ 
                     &&
                     - \left( \delta \mtv{v} \cdot \nabla \Omega \right) r \sin\theta \, \hat{e}_{\phi}
                       \, ,                       \label{eq:dfdt}  \\
\left( \frac{\partial}{\partial t} + \Omega \frac{\partial}{\partial \phi} \right) \delta h  & = & -  \frac{\partial \rho}{\partial h} \, \nabla \cdot  \left( \rho \delta \mtv{v}  \right) \, ,      \label{eq:dPdt} \\
\nabla^2 \delta \Phi & = & 4 \pi G \, \delta \rho \, . \label{eq:dPhi-poiss}
\end{eqnarray}
 In equation (\ref{eq:dfdt}) we have used   the
$\phi$-component of the orthonormal basis vectors; $\hat{e}_{\phi}$.

\begin{table*}
\begin{center}
\caption{\label{tab:back-models} This table lists key
quantities for three sequences of differentially rotating equilibrium
configurations. The stellar models are described by a  $\gamma = 2$ polytropic
equation of state and the j-constant 
rotation law. The first column gives the parameter $A$ that
controls the degree of differential rotation. In the second and third 
columns, we provide, respectively, the ratio of polar to equatorial axes
and the parameter $j_0$ .  The mass,  maximum pressure,  angular momentum and the
rotational parameter 
$\beta = T/|W|$ are given in the last four 
columns, respectively. All quantities are expressed in dimensionless
units, where $G$ is the gravitational constant, $\rho_m$ represents
the maximum mass density and $R_{eq}$ is the equatorial radius.}
\begin{tabular}{c  c c c c c  c }
\hline
 $A/R_{eq} $ &  $ R_p / R_{eq} $  &  $ j_{0} / (\sqrt{G\rho_m} R_{eq}^2 )$ &  $ M / (\rho_m R_{eq}^3)$   
  & $p_m / ( G R_{eq}^2 \rho_m ^2 )  $  & $ J / ( G^{1/2} \rho_m ^{3/2} R_{eq}^5)$  & $ \beta \times 10^{2}$ \\
\hline
 1.0 &  1.0            &         0.000       &      1.273      &   0.637 & 0.000     &   0.00      \\
 1.0 &  0.9            &         0.569       &      1.167      &   0.570  &0.125     &   2.49      \\ 
 1.0 &  0.7            &         0.964       &      0.947      &   0.431  &0.174    &    8.36      \\ 
 1.0 &  0.5            &         1.189       &      0.718      &   0.285  &0.165    &    15.82    \\ 
 1.0 &  0.3            &     	1.337  	  &       0.558 	&  0.149  & 0.152    &     25.57     \\
\\
 0.5 &  0.9            &		0.231	&	 1.229		&   0.585  & 	0.125 	& 2.33  \\
 0.5 &  0.7            &     	0.413  	&  	1.161            &   0.474  &  0.219	& 7.72\\
 0.5 &  0.5            &        	0.547 	& 	 1.116           &   0.359 &   0.294 	& 14.07  \\
 0.5 &  0.3            &       	0.605  	& 	 0.997           &   0.259 &   0.318 	& 20.05    \\
 0.5 &  0.1            &      	0.593 	&  	 0.891           &   0.209  &  0.299 	& 22.43  \\
\\
 0.1 &  0.9          &         	0.049     &      1.329      	& 0.623    	& 0.059  		&   	0.80  \\
 0.1 &  0.7           &         	0.091     &      1.356      	& 0.592  	& 0.114  		&   	2.53  \\
 0.1 &  0.5            &    		0.121 	&  	1.355  		&  0.559 	& 0.152 			& 	4.23    \\
 0.1 &  0.3            &    		0.143 	& 	1.344 		& 0.531 		& 0.531 			& 	5.59     \\
 0.1 &  0.1    		&  		0.154      & 	1.336   		& 0.515  	& 0.193  		&   	6.36  \\
\hline   
\end{tabular}
\end{center}
\end{table*}

To simplify the numerical study of equations (\ref{eq:dfdt})-(\ref{eq:dPhi-poiss}) we 
perform a Fourier expansion of the variables with respect to the $\phi$-angle, leading to the introduction of the azimuthal harmonic index $m$. This way, we only have to 
 solve a  two-dimensional problem in space, rather than the original three-dimensional one~\citep{1980MNRAS.190...43P}. 
For any perturbation variable we use an  expansion, analogous to the following  for the mass density;
\begin{equation}
\delta \rho \left( t,r,\theta,\phi \right) = \sum_{m=0}^{m=\infty}
               \left[ \delta \rho_{m}^{+} \left( t,r,\theta\right)
               \cos m \phi + \delta \rho_{m}^{-} \left(
               t,r,\theta\right) \sin m \phi \right] \, .
               \label{eq:dPexp}
\end{equation}

For any $m$, we  numerically evolve a system of ten partial differential
equations for the twelve variables $\left( \delta \mtv{v}^{\pm}, \delta
h^{\pm}, \delta \Phi^{\pm}, \delta \rho^{\pm} \right)$, while  the density 
perturbation is determined from the EoS.   For our polytropic stars, $\delta \rho$ is simply given by
\begin{equation}
\delta \rho = \frac{\rho}{c_s^2} \delta h \, ,
\end{equation}
where the speed of sound is defined by
\begin{equation}
 c_s^2 = \frac{\partial P }{\partial \rho } = k \gamma \rho ^{\gamma-1}   \, .
\end{equation}

Finally, we introduce the Lagrangian displacement vector $\pmb{\xi}$. This is a key quantity in the instability criteria developed by~\citet{1978ApJ...221..937F}, and we also need it to implement some of the boundary conditions associated with the problem. We determine the Lagrangian displacement by evaluating~\citep{1978ApJ...221..937F}:
\begin{equation}
\delta \mtv{v} = \frac{\partial \pmb{\xi}}{\partial t} + \mtv{v} \cdot \nabla \pmb{\xi} -  \pmb{\xi} \cdot \nabla \mtv{v}  \label{eq:xi} \,  ,
\end{equation}
at each time step during the numerical simulations.

\subsection{Boundary conditions} \label{sec:BC}

The evolution equations need to be complemented by boundary
conditions. At the stellar surface we require that the Lagrangian
perturbation of the enthalpy vanishes, i.e.,
\begin{equation}
\Delta h = \delta h + \pmb{\xi}  \cdot \nabla h = 0  \, . \label{eq:DP-bc}
\end{equation}
We satisfy this boundary condition by imposing the condition 
\begin{equation}
\delta h  =  - \pmb{\xi}  \cdot \nabla h \, ,\label{eq:DP-bc2}
\end{equation}
at the surface.
All  other variables are extrapolated at the surface
grid point at each time step and the gravitational potential is
determined by solving the Poisson equation.

For non-axisymmetric oscillations with $m \geq 2$,  the regular behaviour of
equations~(\ref{eq:dfdt})-(\ref{eq:dPhi-poiss}) at the origin, $r=0$,
and on the rotational axis, $\theta = 0$,
is guaranteed by the conditions:
\begin{equation}
\delta h = \delta \rho = 0 \, ,  \quad \textrm{and} \quad \delta \mtv{v} = \mtv{0} \, .
\end{equation}

Finally, at the equator, $\theta = \pi/2$, the 
perturbation variables divide in two classes with opposite reflection
symmetry. In the first class,  the  variables $\delta \rho^+ , \delta \Phi^+ , \delta v_r^+, \delta v_{\phi
}^+$ are all even under reflection with respect to the equatorial
plane, while $\delta v_{\theta}^- $ is odd.  
By contrast, for the second
class $\delta \rho^- , \delta
\Phi^- , \delta v_r^-, \delta v_{\phi}^-$ are odd and $\delta v_{\theta}^+ $ is even.

\subsection{Differentially rotating models} \label{sec:Stmodel}

We construct models of rapidly and differentially rotating polytropes  by solving equations~(\ref{eq:bgmu_x})-(\ref{eq:Psi}) for a given rotation law. Following much of the relevant literature on the subject, we assume the so-called j-constant rotation law, which is defined by
\begin{equation}
 \Omega = \frac{j_0}{A^2+r^2 \sin^2 \theta} \, ,
 \label{eq:Jlaw}
\end{equation}
where   $j_0$ is a constant  
 and $A$ is a  parameter that controls the degree of differential rotation. 
In the  limit of small $A$,   equation~(\ref{eq:Jlaw}) generates
 a configuration with  constant  specific angular momentum (hence the name of the rotation law), while for  large $A$ 
the star tends to uniform rotation. 
 The constant $j_0$ can be easily obtained by considering equations~(\ref{eq:Jlaw})  on the rotation axis, which leads to 
 $j_0 = \Omega_{c} A^2 $, where $\Omega_{c}$ is the star's angular velocity  on the rotation axis. 

In practice, we solve equations~(\ref{eq:bgmu_x})-(\ref{eq:Psi})  by means of the self-consistent field method described by~\cite{1986ApJS...61..479H}.  
This consists of an iterative numerical procedure,  which can be started by giving as an input  the polytropic index of the EoS, the ratio of the polar to equatorial axes
$R_p/R_{eq}$, and the degree of differential rotation via the parameter $A$. 
Our numerical code for determining differentially rotating stars builds on  
the code developed by~\citet{2002MNRAS.334..933J} to study uniformly rotating configurations. 

In Table~\ref{tab:back-models} we provide data for  
 three sequences of differentially rotating models. The models are described in terms of
dimensionless quantities obtained from the maximum density $\rho_{m}$, equatorial radius
$R_{eq}$ and gravitational constant $G$~\cite[as in for instance][]{1986ApJS...61..479H}. 
Note that, in dimensionless units the differential rotation parameter is given by $\hat A = A/R_{eq}$. 
Nevertheless, we will for clarity use  $A$ instead of $\hat A$ in the discussion of the results.

From the data reported in Table~\ref{tab:back-models}, we
can construct different sequences of rotating stars, eg. with constant
mass or constant angular momentum.  For instance, by specifying the
mass of the star $M$ and the EoS parameters $k$ and $\gamma$, we can
obtain the equatorial radius and the maximum mass density in physical
units from the following expressions:
\begin{eqnarray}
 R_{eq} & = & \left[ \frac{1}{G} \frac{k}{\hat{k}}  \left(\frac{M}{\hat M}\right)^{\gamma-2} \right] ^{1/(3\gamma - 4)}     \, , \\ 
 \rho_{m} & = &  M \hat M ^{-1}  R_{eq}^{-3} \, ,
\end{eqnarray}
where for a polytropic EoS $\hat k$ is equal to the maximum dimensionless pressure $\hat p_{m}$, 
and the  ``hats'' denote the dimensionless quantities listed in Table~\ref{tab:back-models}. 

\subsection{Code description and tests} \label{sec:code}

Having determined suitable background configurations, we set up a numerical code that evolves in time the system of
hyperbolic perturbation equations~(\ref{eq:dfdt})-(\ref{eq:dPdt}), and at the same time
solves (at each time step) the perturbed Poisson
equation~(\ref{eq:dPhi-poiss}). The part of the code that evolves the
hyperbolic equations uses the  technology that was developed in previous
work~\citep{2009MNRAS.394..730P, 2009MNRAS.396..951P}, whereas the
elliptic equation~(\ref{eq:dPhi-poiss}) is solved by means of a spectral
method. The numerical grid is bidimensional and covers  the volume
of star in the region $ 0 \leq r \leq R(\theta) $ and $ 0 \leq \theta
< \pi / 2$. With the definition of a new radial coordinate $x =
r/R(\theta)$, we can associate each grid point with a fluid element of
the star, even when the star is highly deformed by rotation.  The
perturbation variables that obey the hyperbolic differential
equations~(\ref{eq:dfdt})-(\ref{eq:dPdt}) are
discretized on the grid and updated in time with a Mac-Cormack
algorithm. Finally, the numerical simulations are stabilised from high
frequency noise with the implementation of a fourth order
Kreiss-Oliger numerical dissipation $\veps_{\rm D} D_{4} \pmb \xi $, with $\veps_{\rm D} \approx 0.01$. More technical details on the numerical implementation can be found in~\cite{2009MNRAS.394..730P, 2009MNRAS.396..951P}.

\begin{figure*}
\begin{center}
\includegraphics[height=82mm]{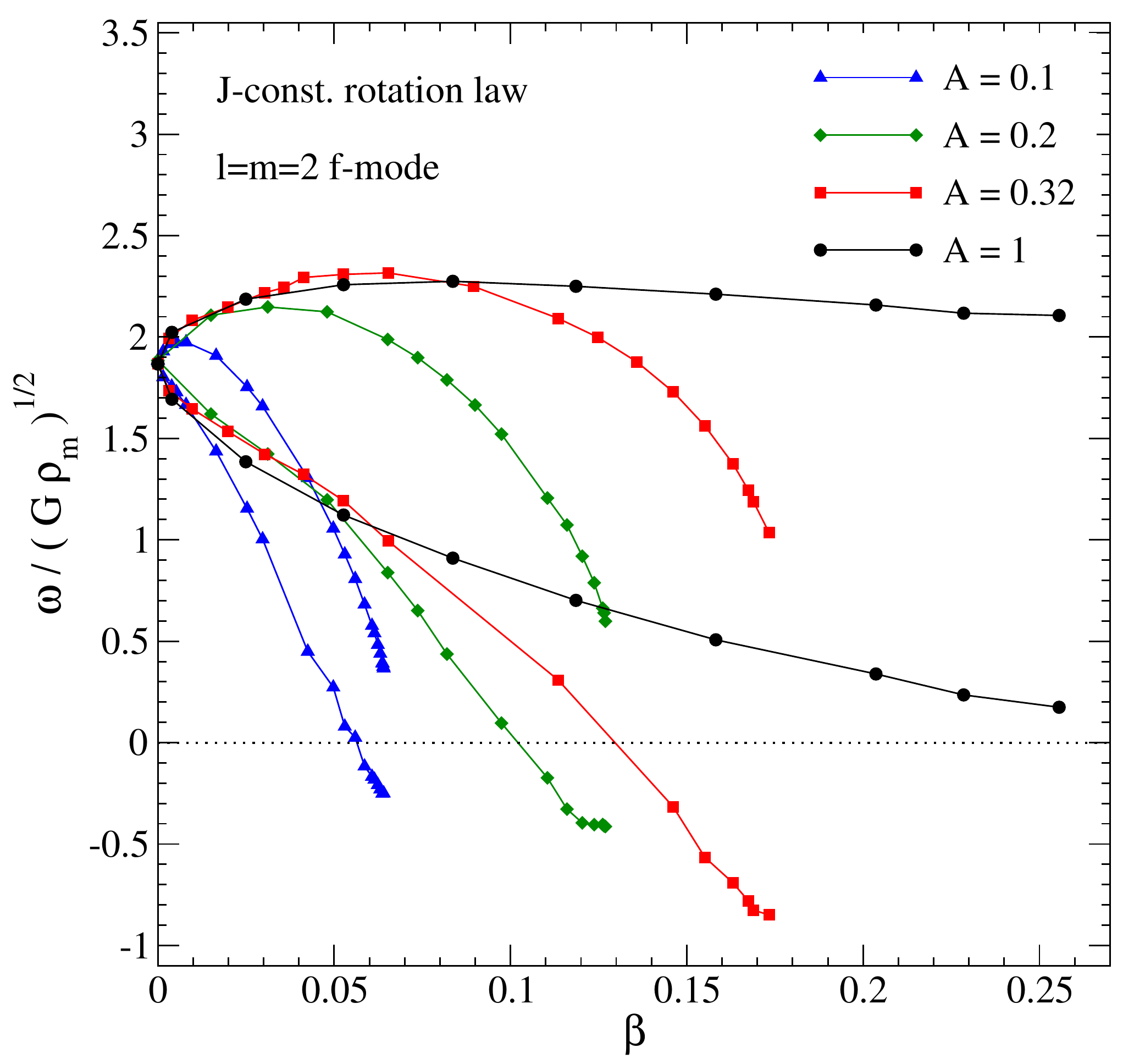} 
\includegraphics[height=82mm]{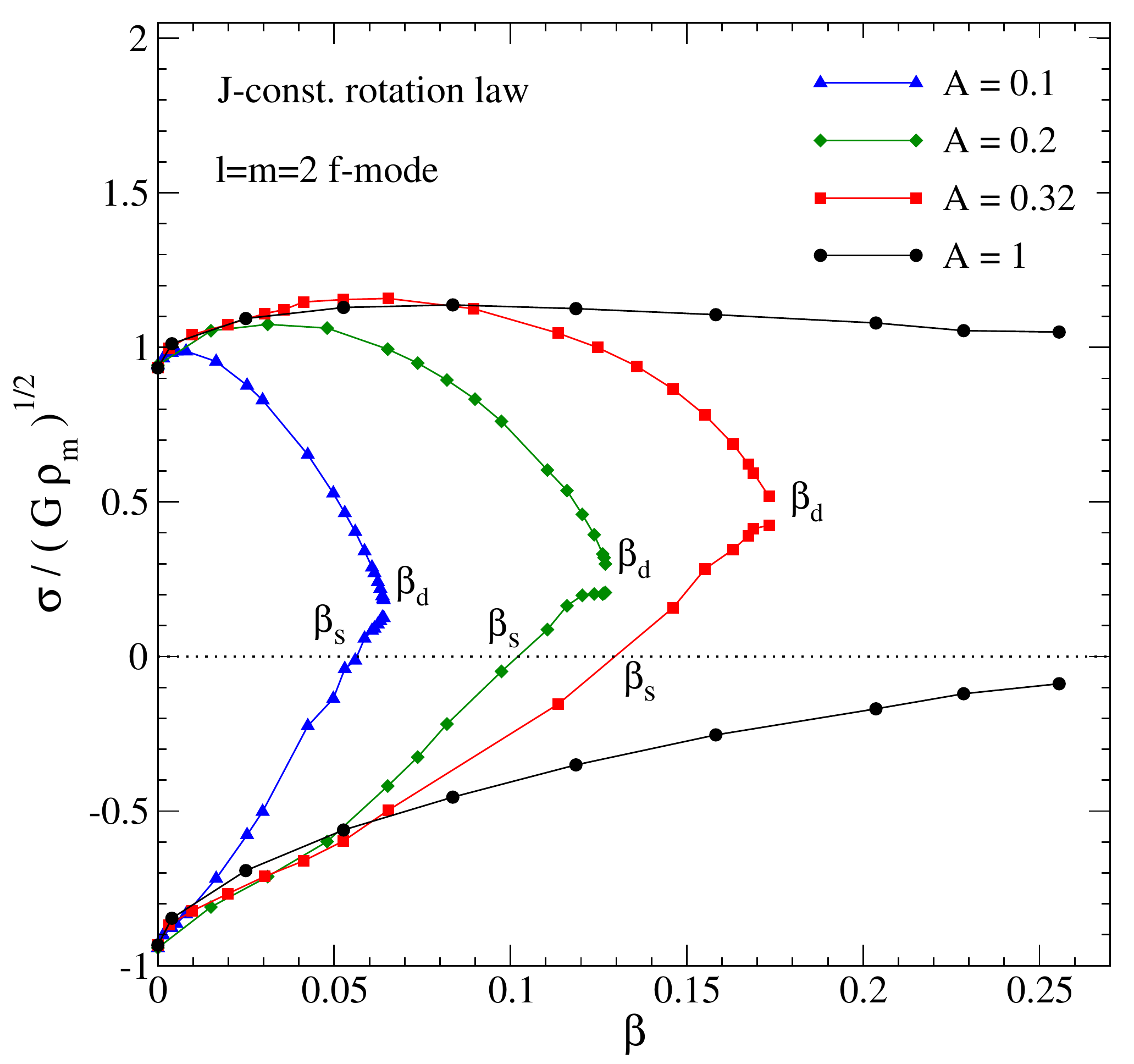} 
\caption{ This figure shows the  $l=m=2$ f-mode frequency (left-hand panel) and the 
pattern speed (right-hand panel) with respect to the star's rotation $\beta = T /|W| $ for four
sequences of differentially rotating stars with, respectively,  $A=0.1,0.2, 0.32$ and
$1$.
The mode frequency and pattern speed are given in dimensionless units.  
The horizontal dotted line in  each panel denotes the neutral line of the secular instability,  $\beta_s$, where an initially retrograde mode (according to an inertial observer) becomes 
prograde as a result of star's rotation. The critical value 
$\beta_d$ (indicated in the right panel) denotes the points in the $\beta-\sigma$ plane where the two branches of $l=m=2$ f-mode 
merge and, consequently, the dynamical bar-mode  instability is expected to develop. 
\label{fig:f22}}
\end{center}
\end{figure*}

Most of the results presented in this paper were obtained using a $48\times90$ grid to cover the $\theta$ and $r$ coordinates, respectively. 
In order to test the accuracy of our growth time extraction we evolved some models with a 
$96\times180$ grid. This showed that the numerical error in the key quantities was significantly less than 1\%, which means that the conclusions we draw from the results should be reliable.

\section{The low T/W instability} \label{sec:LTW}

Our main interest is to trace the onset of the low T/W instability along a given sequence of differentially rotating stars. Specifically, we want to test the assertion of  \cite{2005ApJ...618L..37W}  that the instability is present as soon as the f-mode enters the co-rotation region, i.e. whenever there is a point in the star (the co-rotation point) where the pattern speed of the mode matches the local rotation velocity. That is, when we have
   \begin{equation}
 \sigma = \Omega (\varpi_c) \,  , \label{eq:cor}
\end{equation}
   where $\sigma = \omega / m$ is the pattern speed of the mode and $\varpi_c$ is the co-rotation point. 
 Following the logic of the analogous instability in accretion disks \citep{1984MNRAS.208..721P} one would expect this situation to allow for  an exchange of energy  between the mode and the bulk flow, thus causing the instability. Previous work has not had the ``precision'' to consider this issue in detail. Basically, while full 3D nonlinear simulations have demonstrated that the f-mode has to be well inside the co-rotation region for the instability to be active, they have not resolved the problem near the edge of the co-rotation region. Our perturbative framework is better suited for such an analysis. Of course, we will not be able to make any statements about the nonlinear saturation of the instability or the backreaction on the star, for which a nonlinear analysis is essential.
 
\begin{figure}
\begin{center}
\includegraphics[height=82mm]{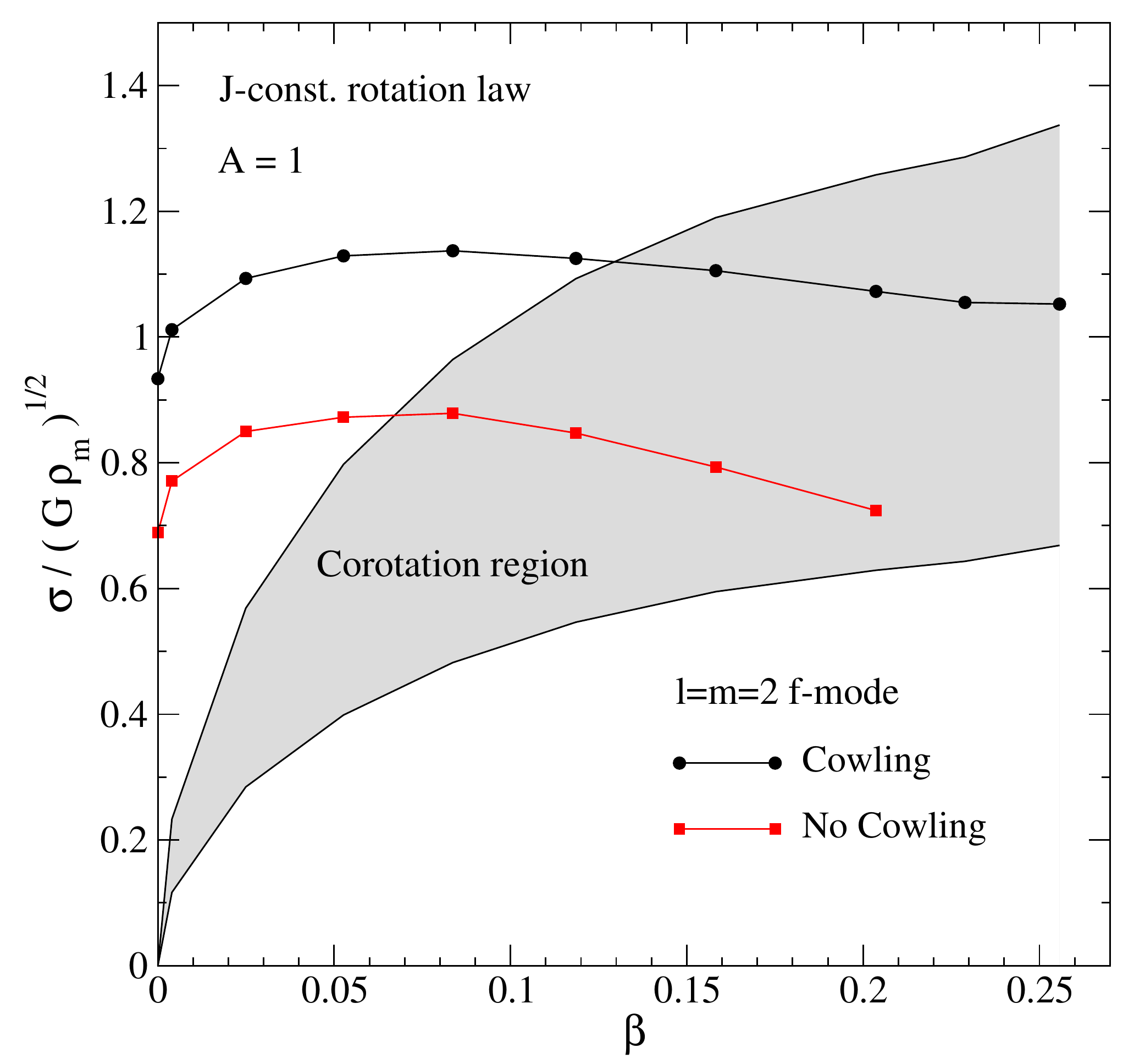}
\caption{The pattern speed, $\sigma = \omega /
m$, of the $l=m=2$ f-mode is shown as a function of the rotation parameter $\beta$ 
for differentially rotating stars with $A=1$. 
The black (dots) and red (squares)  show the values of  $\sigma$ from our simulations, respectively,  in the Cowling and non-Cowling approximation  case.  
The co-rotation band is represented by the grey  region.
\label{fig:cor-A}}
\end{center}
\end{figure}

\begin{figure*}
\begin{center}
\includegraphics[height=82mm]{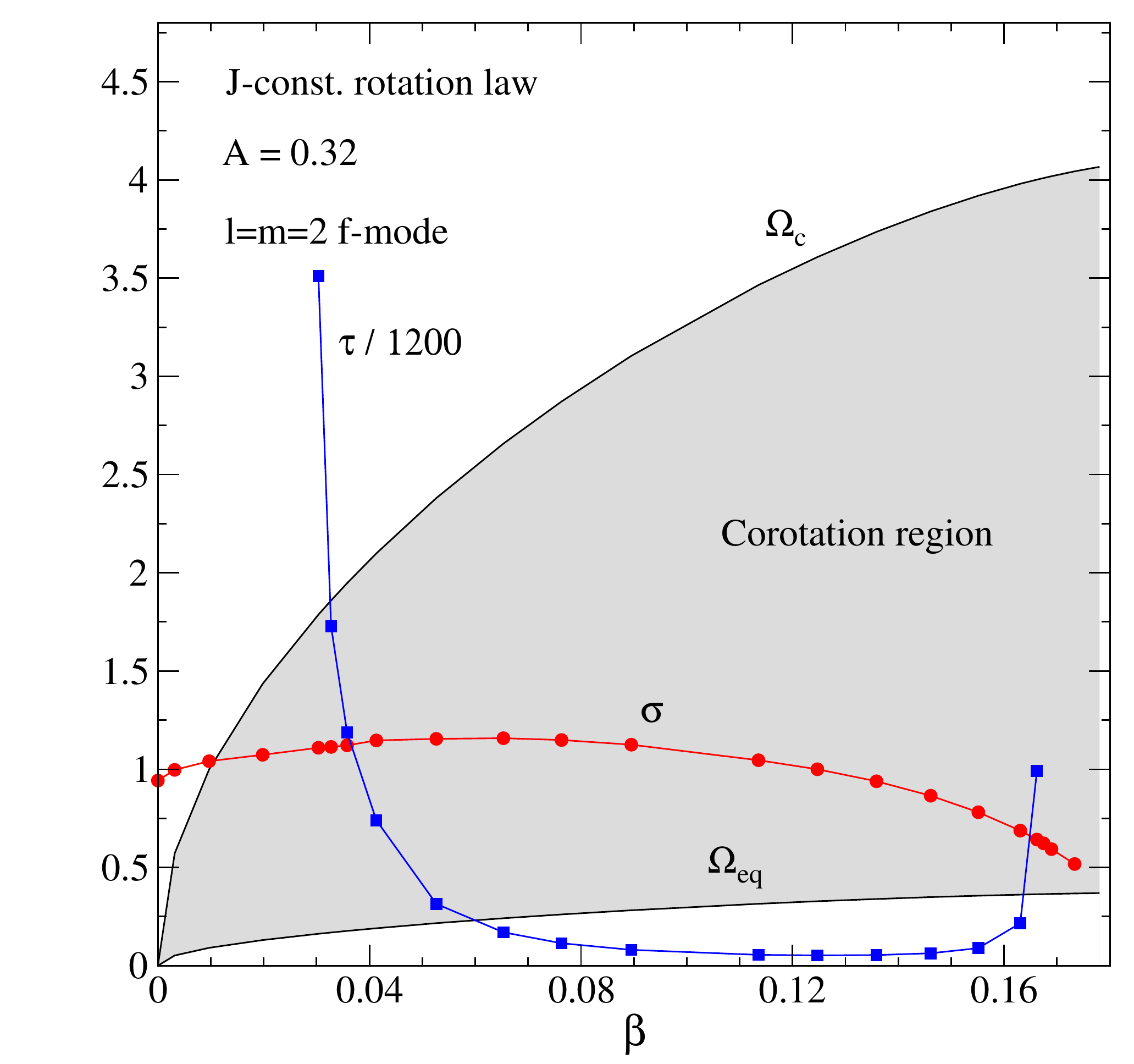}
\includegraphics[height=82mm]{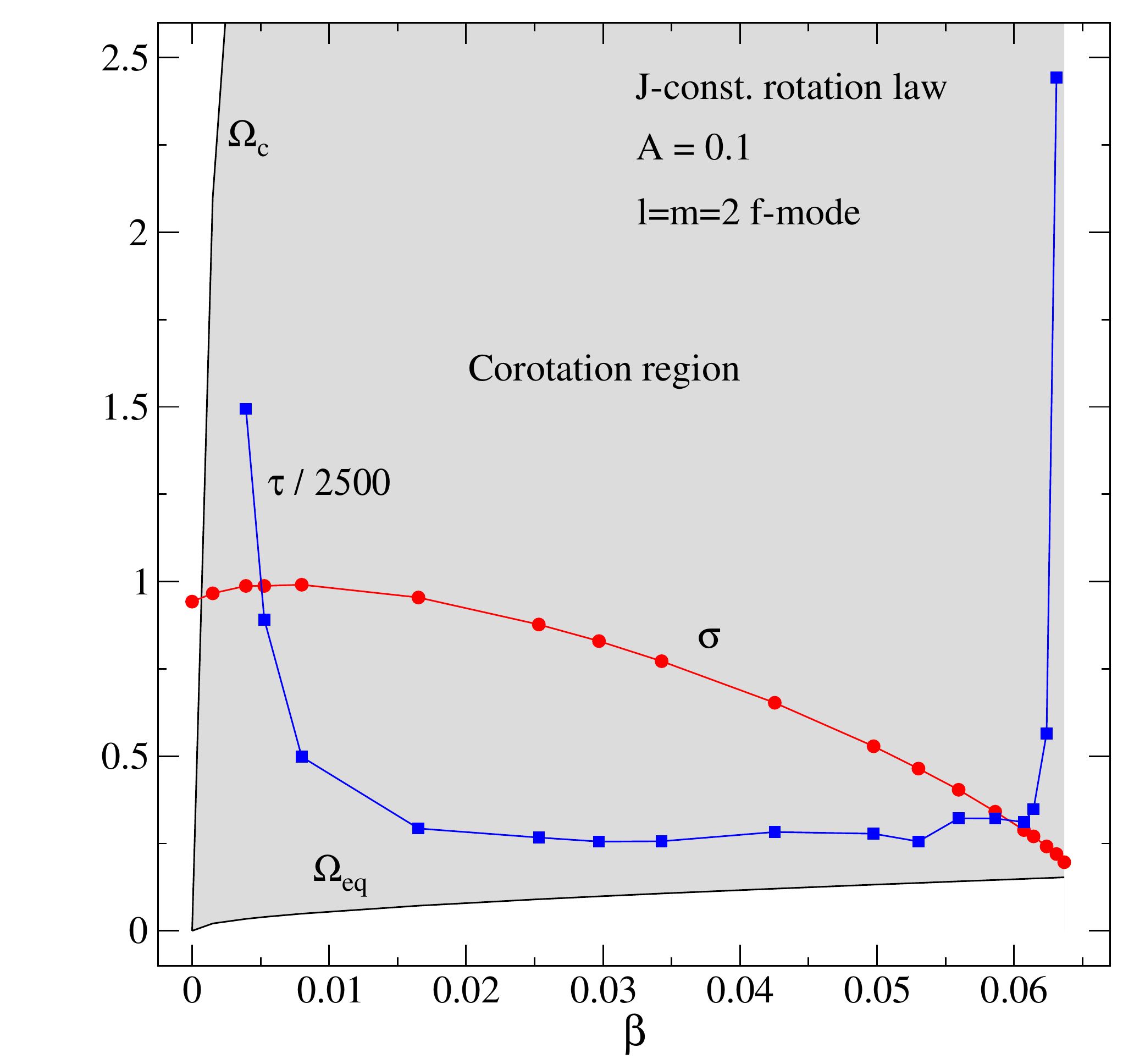}
\caption{This figure provides the inferred shear instability growth time, $\tau$, for differentially rotating  stars 
with $A=0.32$ (left-hand panel) and $A=0.1$ (right-hand panel).  
We also show the f-mode pattern speed, $\sigma $, and the 
co-rotation region (grey).
\label{fig:cor-AB}}
\end{center}
\end{figure*}

 Seminal work on non-axisymmetric instabilities in rotating bodies~\citep{1975ApJ...200..204F, 1978ApJ...221..937F, 1978ApJ...222..281F}  identified  
 key quantities which can be used to monitor the development of an instability;  the canonical energy and angular momentum.     
In an inviscid system a dynamical instability can develop only if these canonical quantities both vanish. This 
guarantees that  an unstable mode does not violate the energy and angular momentum conservation laws. At first, one might think that it ought to be straightforward to use our results to confirm this expectation. Unfortunately, this is not the case. 
First of all, it is not clear that our extraction of the eigenfunctions, which are required for the evaluation of the canonical quantities, is accurate enough due to the growth of the perturbation variables during the evolution.  
Not only are the eigenfunctions  contaminated by numerical noise, there may also be traces of other oscillation modes that are present  during the evolution. 
Secondly, the instability criterion strictly holds only for inviscid systems. In our case, it is not clear what effect the numerical viscosity has on the analysis.  In principle, one could argue that a numerical simulation provides the ``exact'' (inviscid) result as long as the code is convergent. One may, for example, try to extrapolate towards infinite resolution from a set of results with different numerical resolution. Unfortunately, our method is not good enough for us to be able to test this strategy. 
The same is true for the Jordan chain argument from~\citep{1980MNRAS.190...21S, 1980MNRAS.190....7S}, which indicates the degeneracy associated with the merger of two linear perturbation modes (as for example in the case of the classic bar-mode instability at $\beta_d$, see Fig.~\ref{fig:f22}).

Nevertheless, the canonical energy and angular momentum provide important measures. 
In terms of the enthalpy,  the canonical energy  is given by~\citep{1978ApJ...221..937F}: 
\begin{align}
 E_{c}  & = \frac{1}{2} \int d \mtb{r}  \left[  \rho | \partial_{t} \xi_{i} |^{2} - \rho | v^{j} \nabla_{j} \xi_{i}  |^{2} 
+ \rho \, \xi^{i} \xi^{j \ast} \nabla_{i} \nabla_{j} \left( h + \Phi \right)  \right. \nn \\
& \left.+ \frac{\partial h}{\partial \rho} \, | \delta \rho | ^2 
- \frac{1}{4\pi G} | \nabla_{i} \delta \Phi | ^{2}  \right] \, , \label{eq:Ec} 
\end{align}
while the canonical angular momentum follows from~\citep{1978ApJ...221..937F}: 
\begin{equation}
J_{c} = - Re \int d \mtb{r}  \rho \, \partial_{\phi} \xi^{i \ast} \left( \partial_{t} \xi_{i} +  v^{j} \nabla_{j} \xi_{i}  \right)  , \label{eq:Jc} 
\end{equation}
where the integrals are calculated over the star's volume, and  
 $Re$ denotes the real part.  Note that these expressions are given in a coordinate basis, not the orthonormal basis used elsewhere in the paper.
Building on the work of \citet{2006MNRAS.368.1429S} we will use the integrands from  equations~(\ref{eq:Ec}) and~(\ref{eq:Jc})
to  identify the region in the star where the instability develops.

\subsection{The f-mode}

As we have already mentioned, the current understanding is that the low T/W instability sets in 
when the f-mode enters the co-rotation region. In order to support this notion, it is natural to 
determine the mode frequencies along a differential rotation sequence and at the same time keep track of 
 the co-rotation region.   The latter task is straightforward since it only requires the background configurations. The determination of the oscillation frequencies 
for a set of stars along a given sequence is more time consuming. We obtain the required results by first evolving in time 
 the  linearised dynamical equations~(\ref{eq:dfdt})-(\ref{eq:dPhi-poiss}), and then calculating the mode frequencies by performing 
 a Fast Fourier Transformation (FFT)  on the time evolution data.  In order to test the reliability of the method we have confirmed that the results agree with data 
 available in the literature~\citep{PhysRevD.64.024003, 2003MNRAS.343..175K}.

Fig.~\ref{fig:f22} shows the $l=m=2$ f-mode frequencies, measured in the inertial reference frame, 
for four sequences of models with a different degree of differential rotation. 
The stars with the lowest level of differential rotation are parameterised 
by $A=1$, while the highest differential rotation models have $A=0.1$.   For  small values of $A$ the fastest spinning stellar models assume a ``toroidal-like configuration''  
with a small axis ratio and a mass density whose maximum is not at the rotation axis. 
For instance, for $A=0.1$ the fastest model studied in this work has  $\beta = 0.064$ and 
$R_{p}/R_{eq} =0.067$. 

As expected, the results show that 
the f-mode is rotationally split  into two branches  which are, respectively, 
 prograde and retrograde with respect to the star's rotation. 
 The prograde f-mode frequency gradually decreases for higher differential and more rapidly rotating models. Meanwhile, 
 the retrograde f-mode  has an oscillation frequency which decreases with rotation and  can become 
 negative for rapidly rotating models. The neutral point, where the f-mode frequency is zero (indicated by $\beta_s$ in Fig.~\ref{fig:f22}),  is  important as it marks the point 
  where  an f-mode is driven unstable by gravitational radiation via the well-known 
    Chandrasekhar-Friedman-Schutz (CFS) mechanism~\citep{1970PhRvL..24..611C, 1975ApJ...200..204F, 1978ApJ...221..937F}. 
  This secular instability occurs when a locally retrograde mode is dragged forward by the star's rotation to the point where it is seen to be prograde by an inertial observer.   
   
\begin{figure}
\begin{center}
\includegraphics[height=82mm]{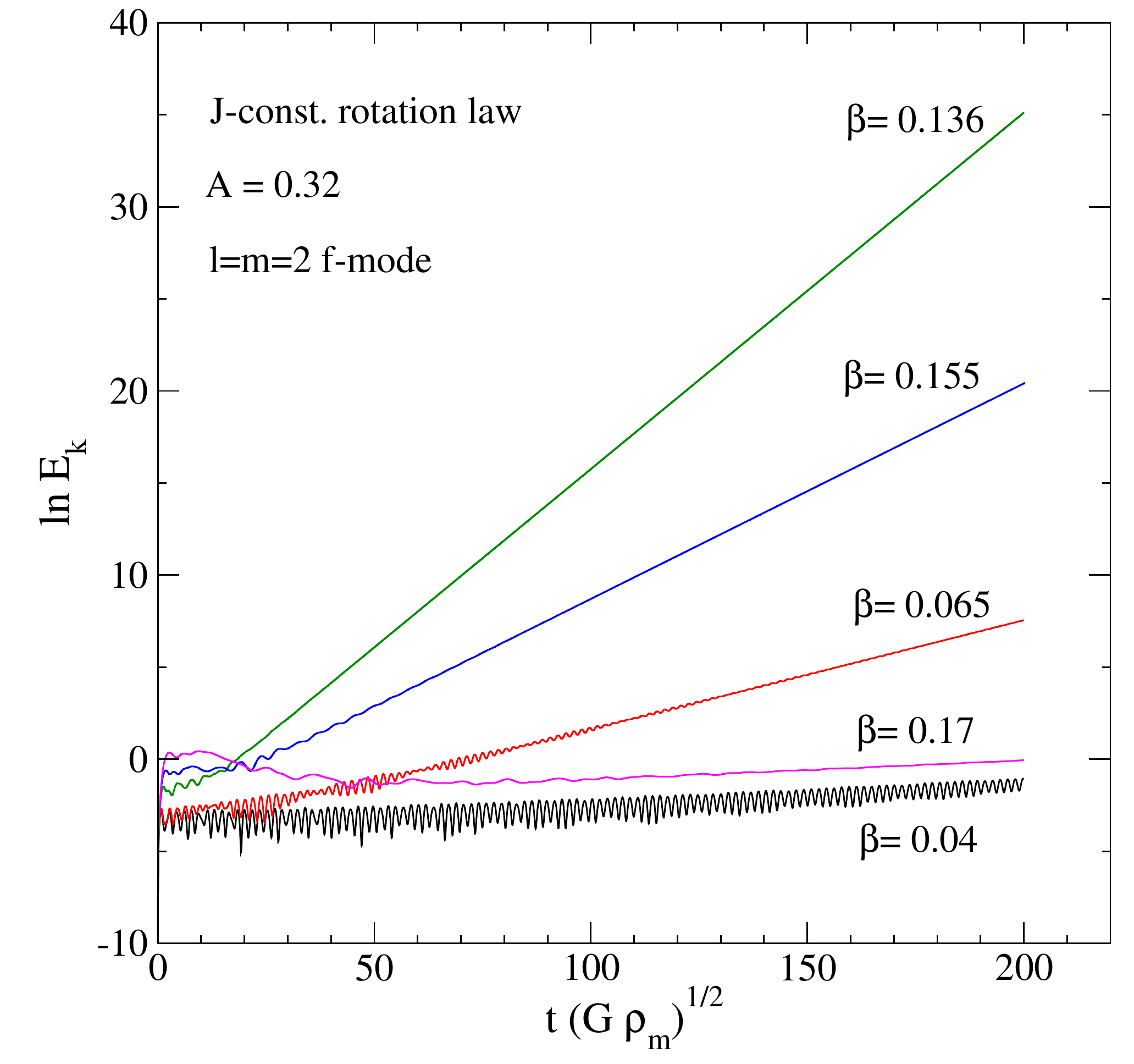}
\caption{A sample of results for the time evolution of the kinetic energy associated with the f-mode for  
models with $A=0.32$ and different values for the rotation parameter $\beta$ (see legend).  The 
evolution time and the kinetic energy are given in dimensionless units. 
The vertical axis displays the natural logarithm of the kinetic energy, $\ln E_k$.
\label{fig:instex}}
\end{center}
\end{figure}

  The conditions for the classical non-axisymmetric instabilities are better illustrated in the right-hand panel of Fig.~\ref{fig:f22}, 
  which shows the pattern speed, $\sigma$, for the $l=m=2$ f-mode against the rotation parameter $\beta$. The results in the figure can be compared to the  results for incompressible 
  ellipsoids \cite[cf. Fig.~5 in][]{2003CQGra..20R.105A}.
  When a mode has a negative (positive) pattern speed,  it is retrograde (prograde) with respect to the star.  
   For the models shown in Fig.~\ref{fig:f22} with $A\leq 0.32$, it is clear that the f-mode is retrograde for slowly rotating stars, but becomes  
   prograde at the critical rotation rate corresponding to $\beta_{s}$. 
   Another significative point in the $\beta-\sigma$ parameter space is where the pattern speeds of the two branches of the f-mode 
   merge. This point indicates where the modes merge and the classic dynamical bar-mode instability sets in in an inviscid star.  Using our perturbative time evolutions we are able to extract 
   the mode frequencies very close to this point, $\beta_{d}$, for  the $A=0.1, 0.2$ and 0.32 sequences. 
   
\begin{figure}
\begin{center}
\includegraphics[height=35mm]{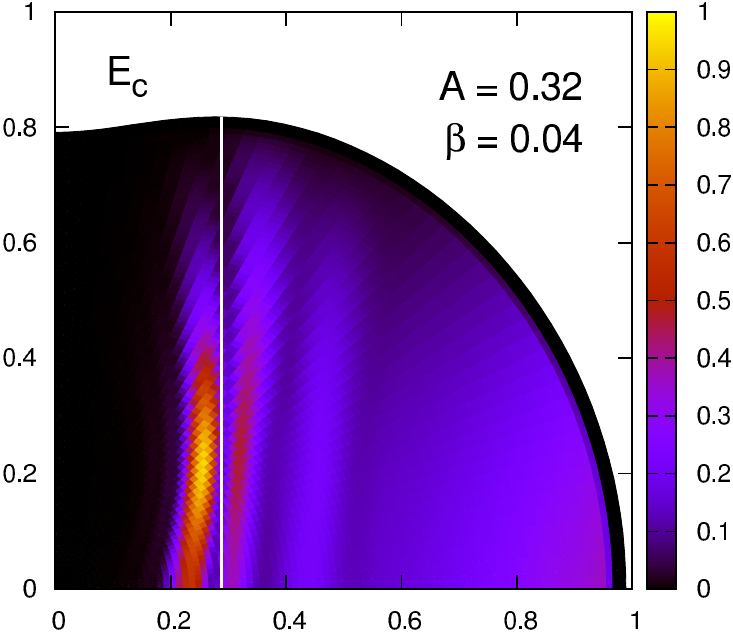}
\includegraphics[height=35mm]{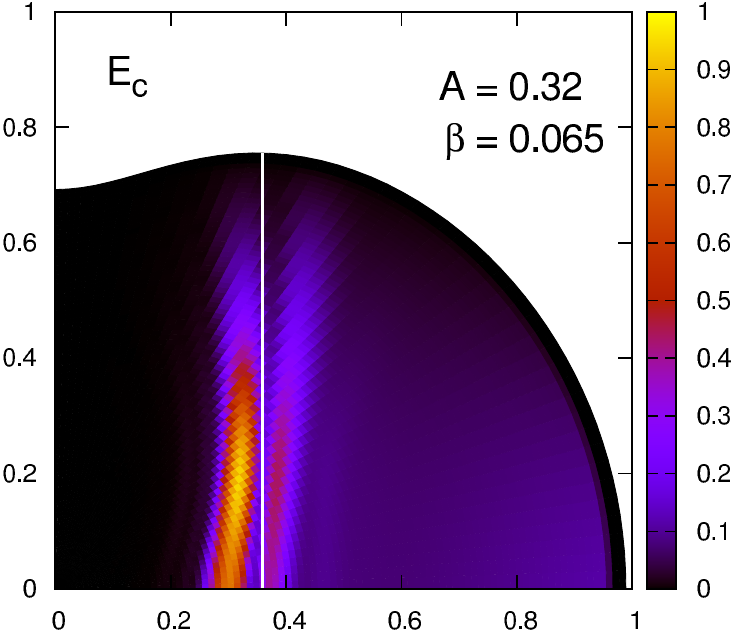}
\includegraphics[height=35mm]{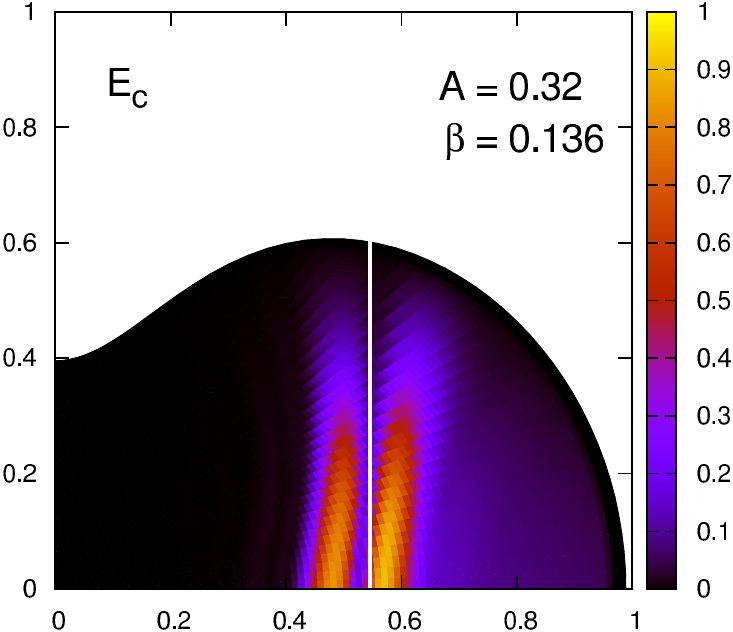}
\includegraphics[height=35mm]{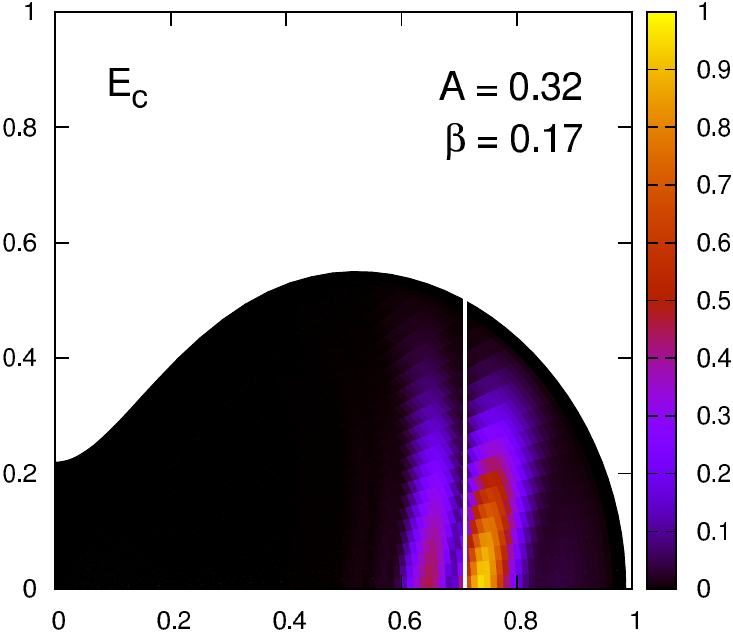}
\caption{Examples of the extracted canonical energy density inside the neutron star. The vertical white line in each panel denotes the 
co-rotation cylindrical radius $\varpi_{cor}$.
\label{fig:Ec}}
\end{center}
\end{figure}

    The part of the parameter space where the low T/W instability might occur is  shown  in Fig.~\ref{fig:cor-A}
     for stellar models with  $A=1$   and in Fig.~\ref{fig:cor-AB} for $A = 0.32$ and $0.1$. 
     For high differential rotation  (small $A$), the co-rotation region increases considerably and the 
    $l=m=2$ f-mode can enter  this  potential instability zone even for very slowly rotating stars. 
  In  Fig.~\ref{fig:cor-A}, we also show the f-mode frequency calculated with the full perturbation problem, i.e. when  the Poisson equation for the gravitational potential is solved alongside the fluid equations.  It is well known  that the Cowling approximation (where the gravitational potential perturbation is neglected) may introduce 
  a significant error in the f-mode frequency~\citep{2003MNRAS.343..175K}. 
   As we can see from Fig.~\ref{fig:cor-A},  the f-mode frequencies are generally smaller when $\delta \Phi$ is included. In the present context this is important since it affect the relation between the mode-frequency and the co-rotation band. As indicated in the Figure, the inclusion of the gravitational potential means that the
    instability may be triggered at lower rotation rates. Unfortunately, 
   the solution of the Poisson equation (an elliptic equation), slows down  
   the numerical simulations significantly which increases the time it would take to perform a large parameter space analysis.  
   In addition, for differentially rotating models our code is more stable when the Cowling approximation is used.  Since we do not expect the Cowling approximation to have any effect on the qualitative behaviour of the low T/W instability we will from now on focus on results obtained by neglecting $\delta \Phi$.

\begin{figure*}
\begin{center}
\includegraphics[height=82mm]{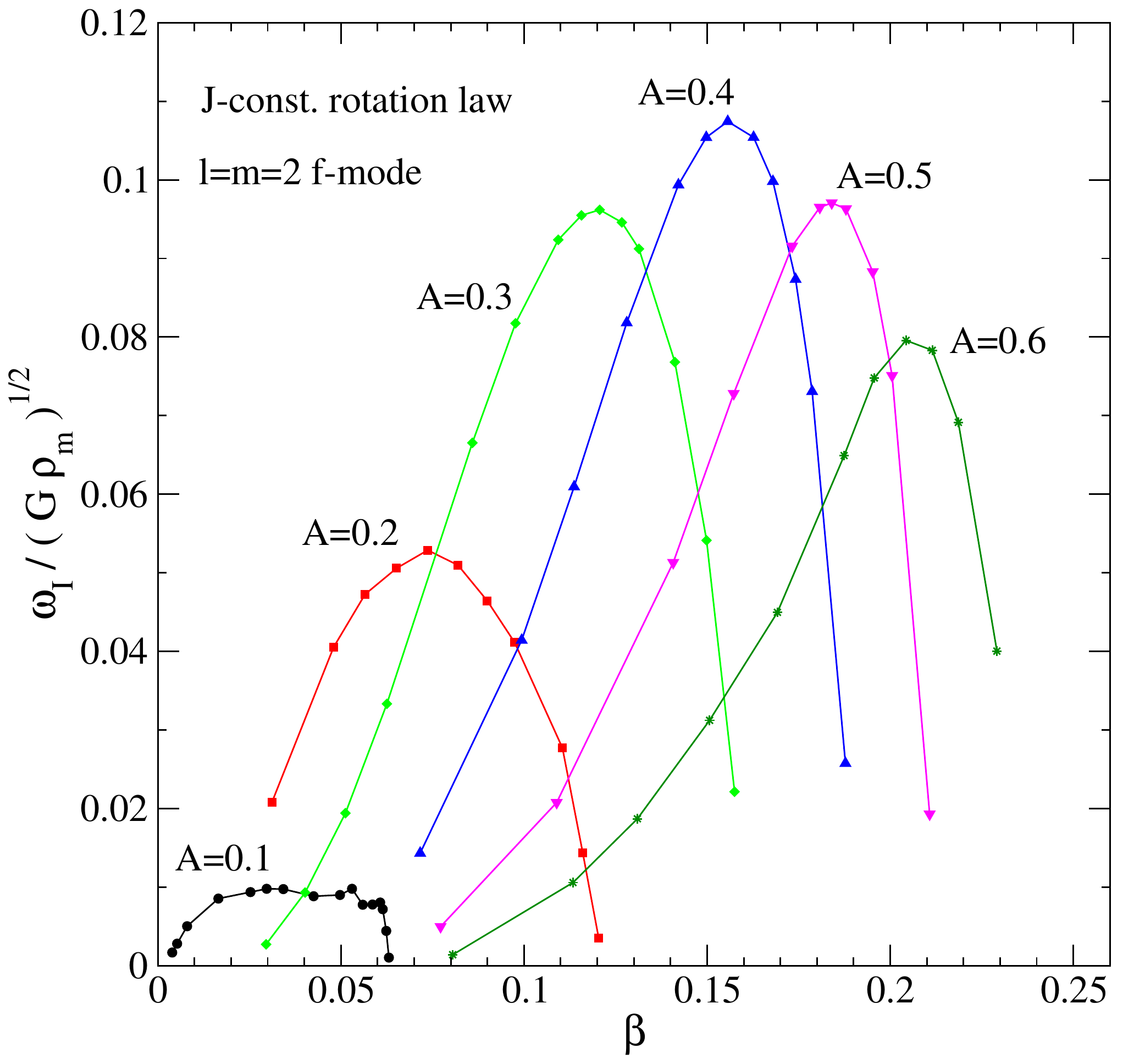} 
\includegraphics[height=82mm]{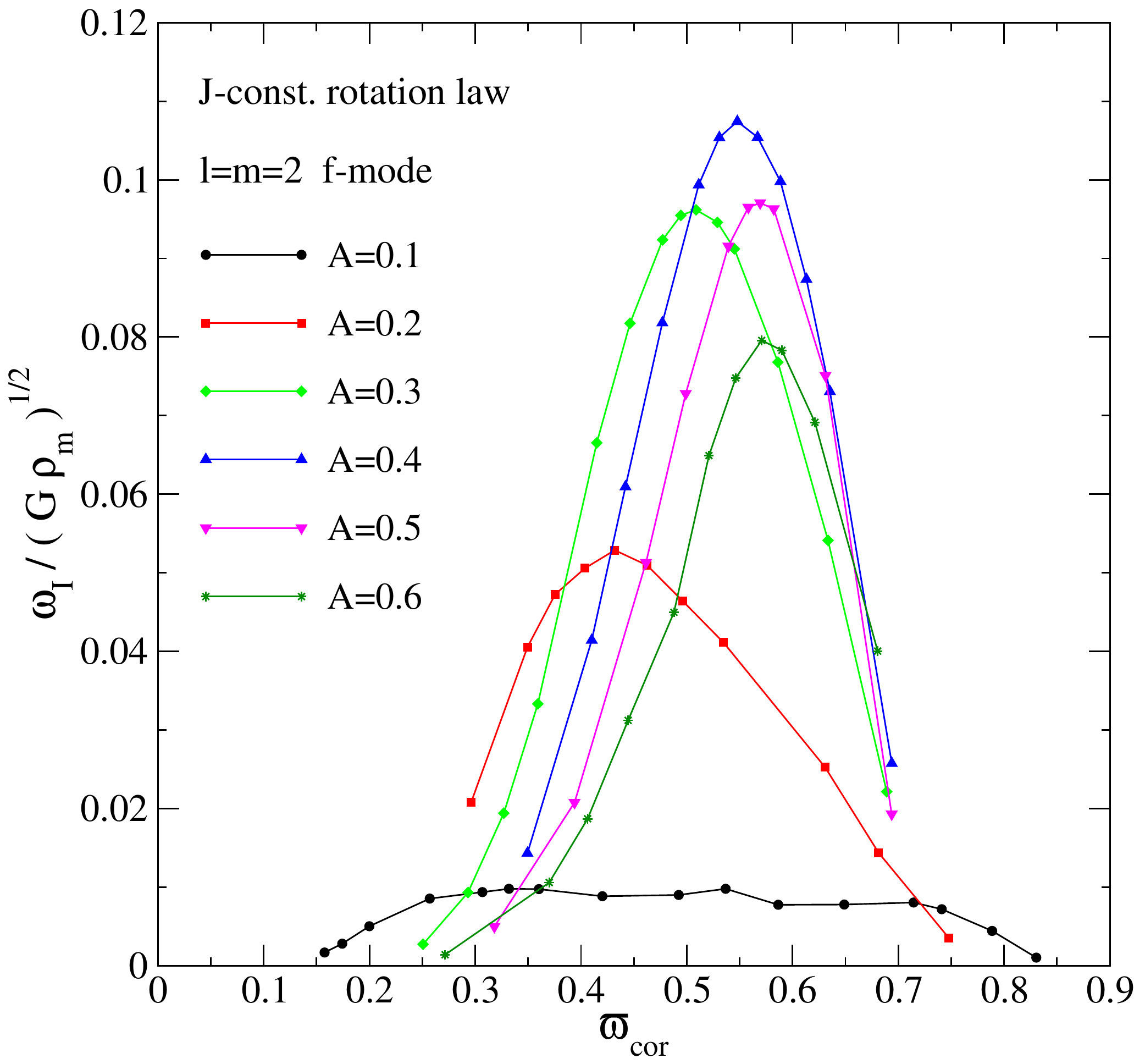} 
\caption{ Imaginary part of the $l=m=2$ mode frequency, $\omega_I = 2 \pi / \tau $,  
for several sequences of differentially rotating stars with $A=0.1-0.6$. The left- and right-hand panels 
show the dependence of $\omega_I$, respectively,  on the rotational parameter $\beta$ and 
the cylindrical co-rotation radius $\varpi_{cor}$. The quantities on the horizontal and vertical axes are given in dimensionless units.
\label{fig:tauA}}
\end{center}
\end{figure*}

\subsection{Instability growth time} \label{sec:Engc}

If the hypothesis from \citet{2005ApJ...618L..37W} is correct, then the  growth rate of the instability should depend on the position of the mode relative to the boundary of the co-rotation region, being more rapid deep inside the co-rotation region and essentially vanish at the co-rotation boundaries (as the mode stabilizes).

In order to  assess this idea we determine  the instability growth timescale from our numerical simulations. As an example, 
we show in Fig.~\ref{fig:instex} the mode kinetic energy, $E_k$, for the rotating star models  with $A=0.32$.  
After some evolution time, the kinetic energy starts to grow exponentially with a rate that 
depends on the stellar rotation. We  determine the instability growth time, $ \tau$, from a linear fit 
to the time evolved kinetic energy by assuming that 
\begin{equation}
E_k \sim e^{ 2  \omega_I t } \, , \qquad  \textrm{where} \quad \omega_I =   \frac{2 {\rm \pi}}{\tau} \, .
\end{equation}
The kinetic energy is defined by the following equation:
\begin{equation}
 E_{k}   = \frac{1}{2} \int d \mtb{r}  \rho \,  \delta \mtv{v} ^{2}  \, .
\end{equation}
For this specific model, we find that the instability timescale initially decreases as the rotation increases 
from a model with $\beta = 0.04$ up to a model with $\beta = 0.136 $,  and then increases again for
more rapidly rotating configurations, i.e. $\beta = 0.17$. 
The rescaled growth times for the $A=0.32$ and 0.1 cases are shown in Fig.~\ref{fig:cor-AB}, together with the pattern speed and the co-rotation region. The results clearly show that that $\tau$ increases towards the boundaries of the co-rotation region. 
In fact, for models very close to these boundaries our time evolutions do not exhibit any instability.
 Most likely this is due to the  numerical viscosity in the code preventing the instability from setting in until the physical growth overcomes the artificial damping. Real viscosity would, of course, play a similar role but it is orders of magnitude weaker than the numerical viscosity used in our simulations.

\begin{figure*}
\begin{center}
\includegraphics[height=50mm]{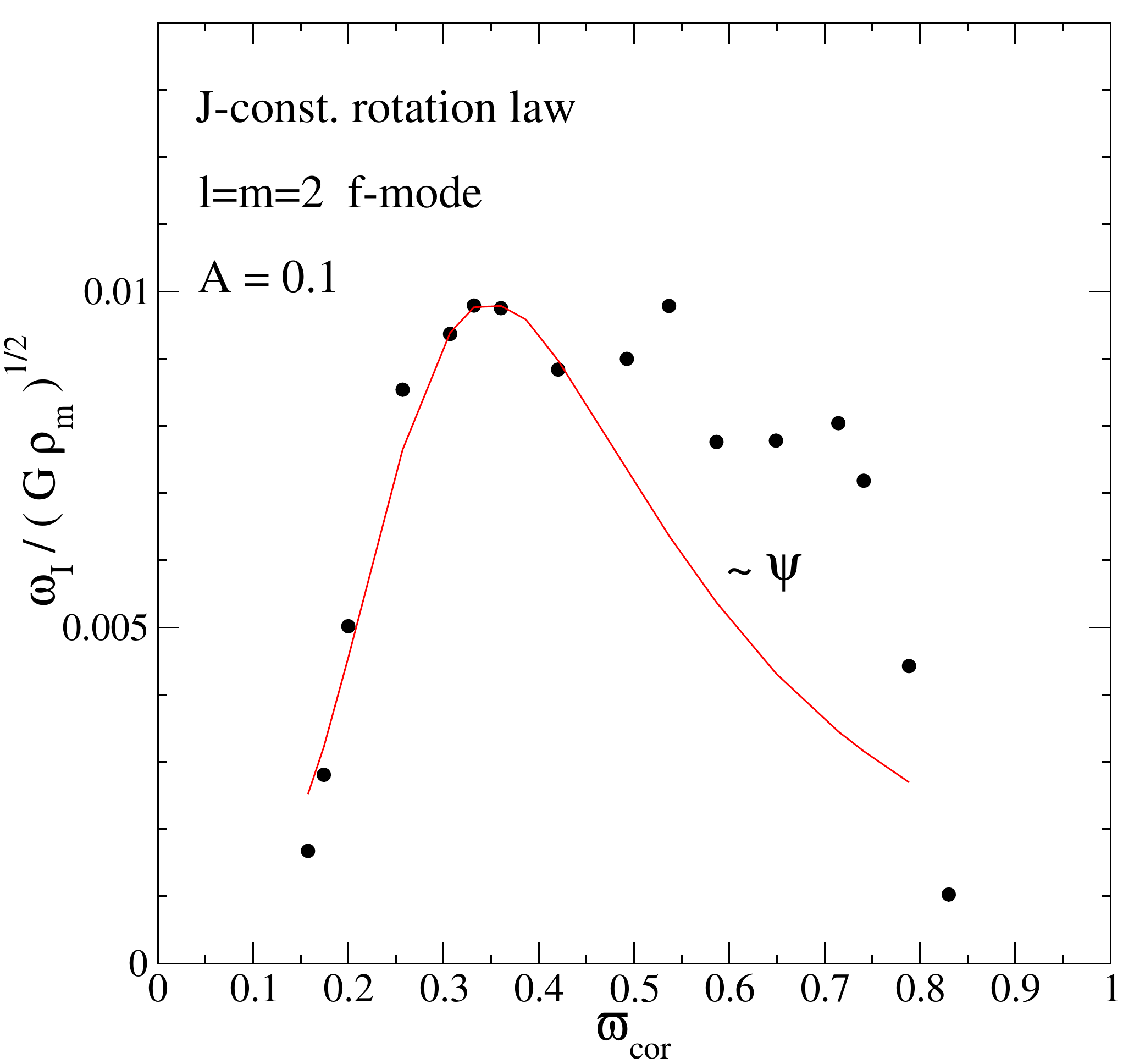} 
\includegraphics[height=50mm]{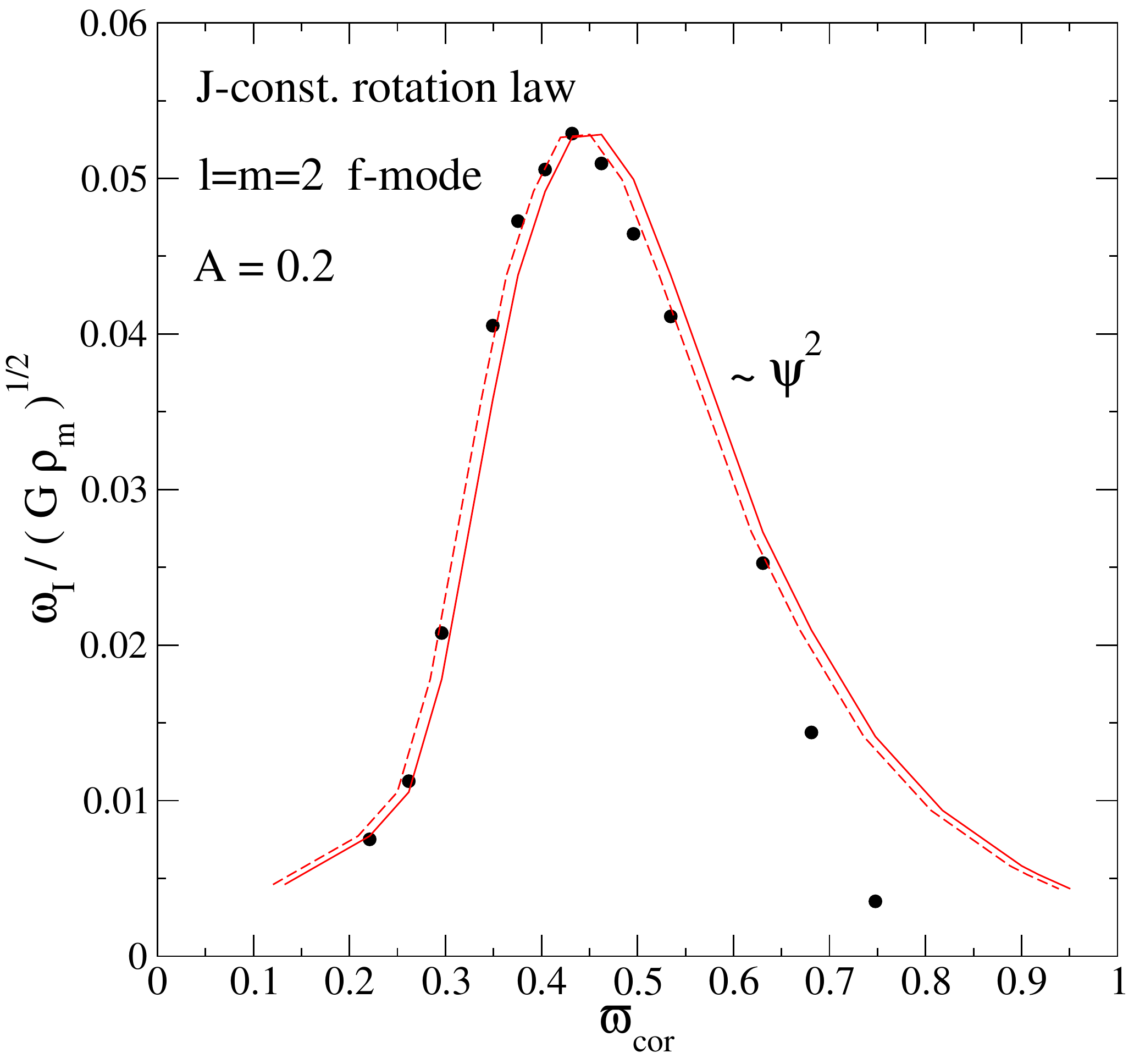} 
\includegraphics[height=50mm]{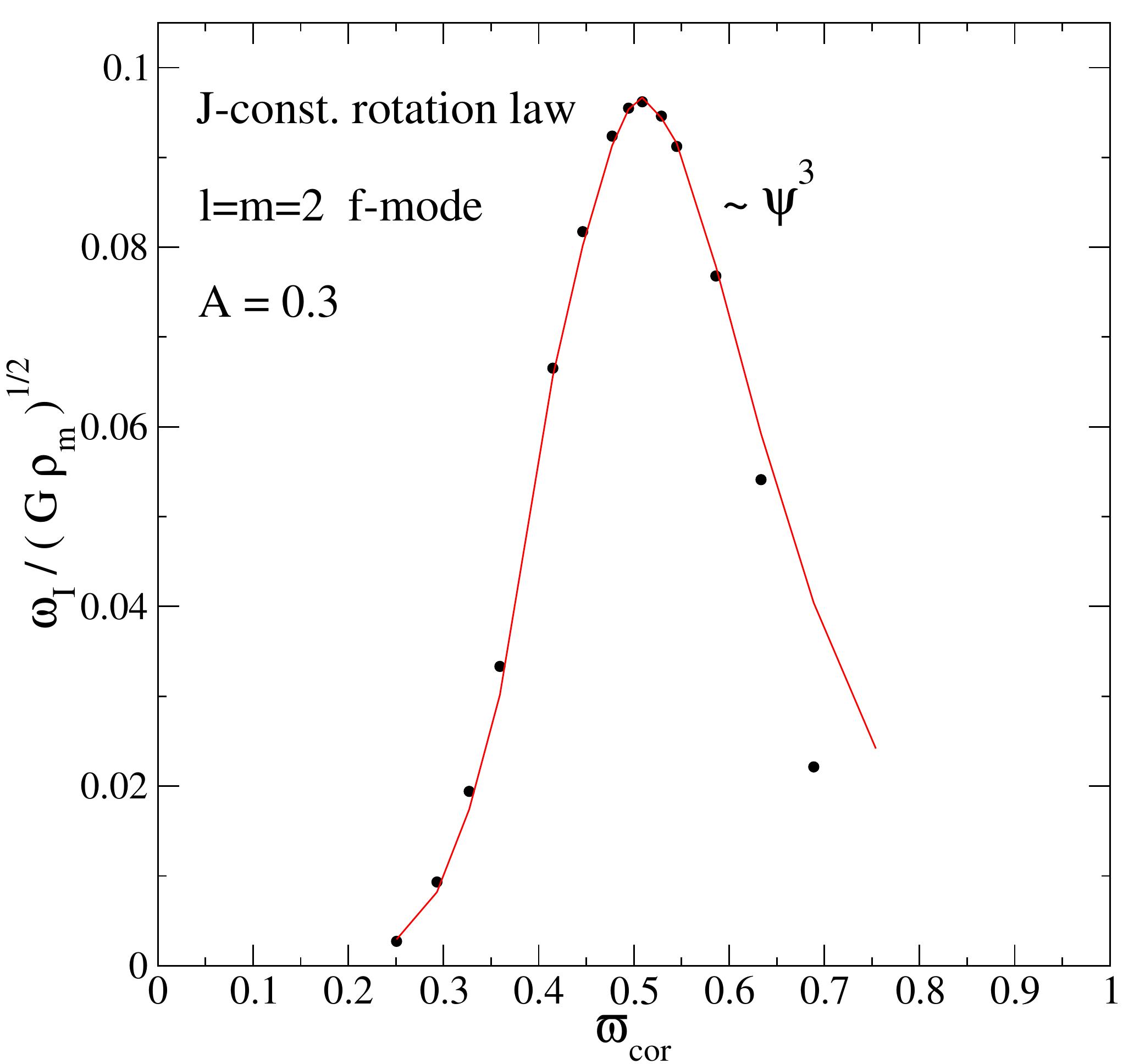} 
\includegraphics[height=50mm]{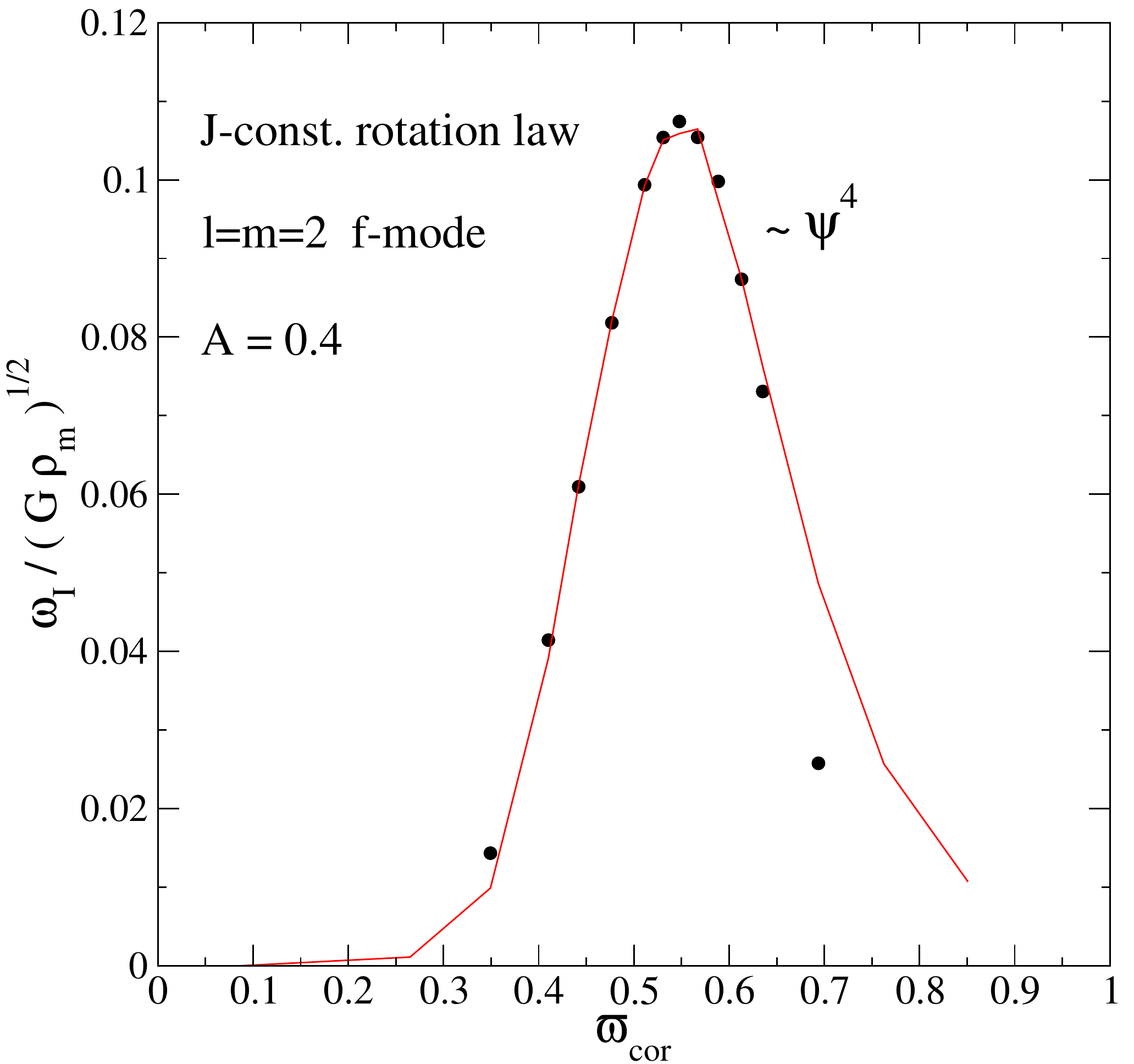} 
\includegraphics[height=50mm]{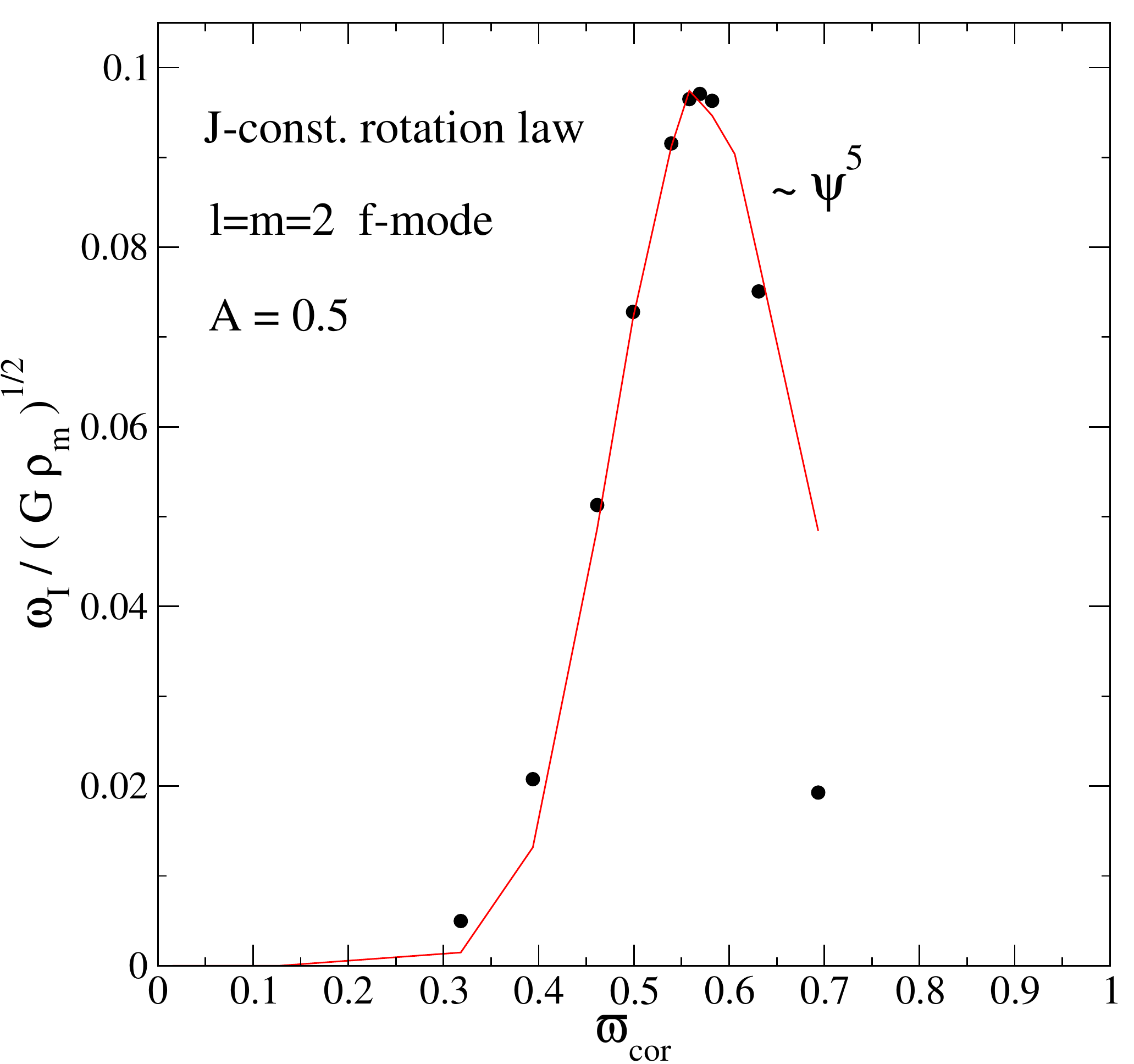} 
\includegraphics[height=50mm]{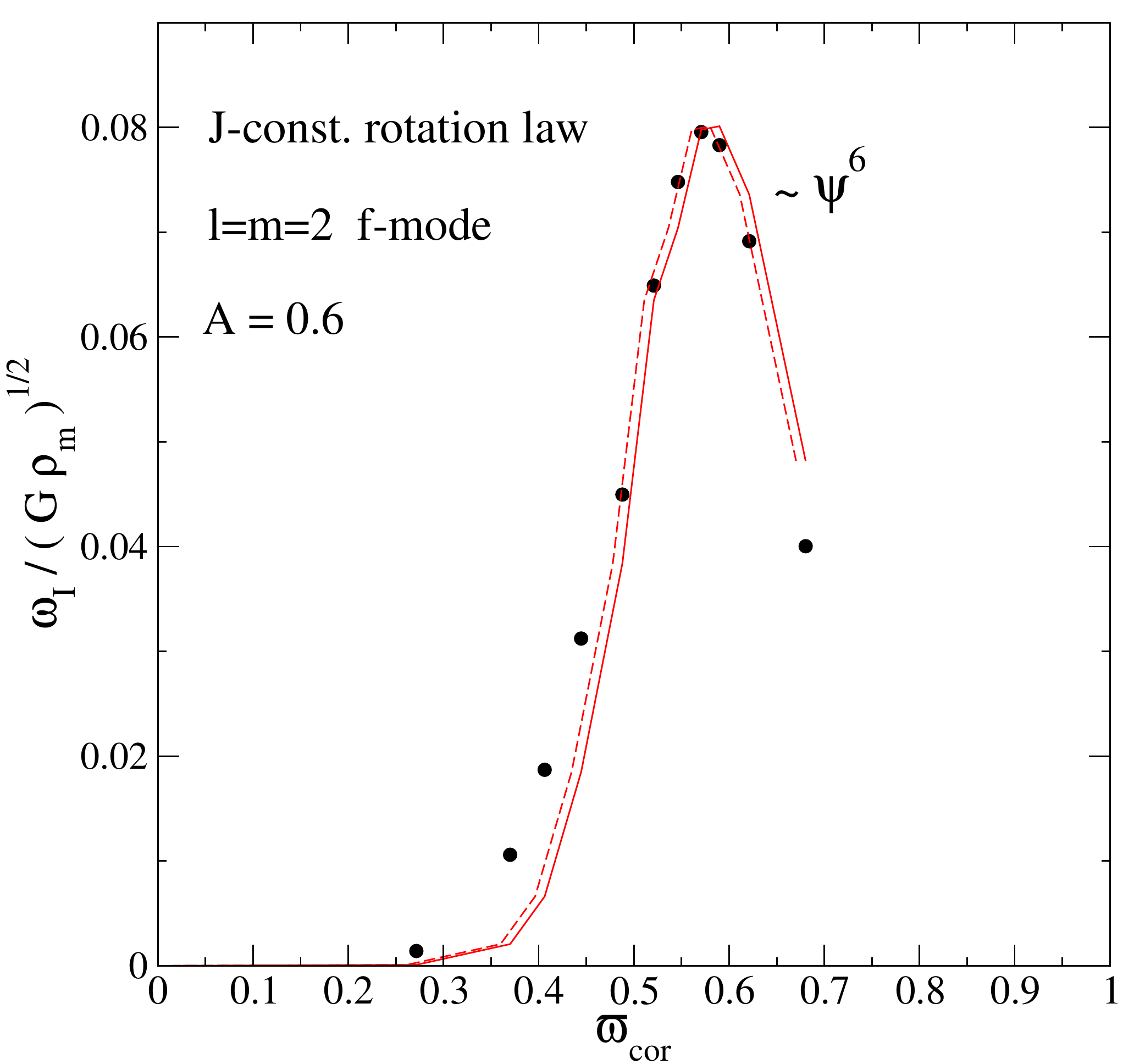} 
\caption{Comparison between the imaginary part of the $l=m=2$ f-mode and the empirical function $\chi$ for six differentially rotating models. 
Each panel corresponds to a model with a specific $A$ (see legend). 
The horizontal and vertical axes show, respectively,   the dimensionless $\varpi_{cor}$ and $\omega_I$. 
The filled dots represent the numerical values and the red lines the function $\chi$. In the panels for the $A=0.2$ and 0.6 models 
we also show a dashed red line, which provides a correction of the function $\chi$ to match the $\omega_I$ values. 
\label{fig:psiA} }
\end{center}
\end{figure*}

In order  to test the effect of the artificial viscosity on the inferred instability growth time we have considered several simulations 
with different numerical dissipation strength for a given stellar configuration. Thus we found that $\tau$ is affected 
by the Kreiss-Oliger dissipation only at the level of about 0.1\% even when the viscosity factor is an order of magnitude larger than the fiducial one used, $\eps_D = 0.01$. Hence, the extracted growth times are not significantly affected by the artificial damping. Of course, it is still the case that the numerical damping will overcome the growth of a very slowly growing unstable mode, e.g. near the co-rotation boundary.

In order to  check that we are tracking a given unstable mode along the stellar sequence, we monitor  the canonical energy and angular momentum 
integrands (see equations~\ref{eq:Ec}-\ref{eq:Jc}). Qualitatively, one would expect the unstable mode to be associated with the region near the  co-rotation point in the star. If the low T/W instability is the result of an energy exchange between a co-rotating mode and the fluid bulk flow, this should be reflected in the energy and angular momentum integrands,. This behaviour was, in fact,  observed for the canonical angular momentum by~\citet{2006MNRAS.368.1429S}, who  showed that  the $J_{c}$ integrand grows when an instability is present while remaining zero at the co-rotation point. 

To study the behaviour of $E_c$ and $J_c$  it is  useful to determine the cylindrical radius, $\varpi_{cor}$, 
at which a given mode co-rotates with the star (where 
$ \varpi = r \sin \theta$). For the j-constant rotation law, we can determine from equation~(\ref{eq:Jlaw}) and the mode frequency, $\omega$,  the following  expression for the co-rotation point:
\begin{align}
& \varpi_{cor} = A \sqrt{  \frac{ m \Omega_c  - Re\ \omega}{Re\ \omega} }  \label{eq:wc} \, .
\end{align}

In Fig.~\ref{fig:Ec} we show the canonical energy  
distribution inside the star and the co-rotation radius $\varpi_{cor}$ calculated from equation~(\ref{eq:wc}),
for a selection of unstable models with $A=0.32$.
As expected, $E_c$ grows around the co-rotation point and vanishes at $\varpi_{cor}$.  The same feature has been observed 
for the canonical angular momentum. 
These results confirm that the growth time extracted from our numerical simulations is  associated with  the $l=m=2$ f-mode (as expected).  

The dependence of the instability growth time on the position of the co-rotation point is evident from Fig.~\ref{fig:cor-A}.    
In order to understand the behaviour better, we consider the variation of $\tau$ with the degree of differential rotation and try to 
relate the results with the stellar parameters.   Typical results are provided in Fig.~\ref{fig:tauA}, where we show
the variation of the unstable mode's imaginary part, $\omega_I = 2 \pi / \tau $,  with respect to the rotation parameter $\beta$ (left-hand panel) 
and cylindrical co-rotation radius (right-hand panel) for stars with $A$ ranging between 0.1 and 0.6. Note that $\varpi_{cor}=0$ corresponds to the 
the condition $\sigma=\Omega_{c}$, while $\varpi_{cor}=1$ implies that $\sigma=\Omega_{eq}$. 
Along each rotating sequence, corresponding to a specific $A$, $\omega_I$ exhibits a maximum inside the co-rotation band and decays towards the boundaries of the region. It is also interesting to note  that for higher differential rotation the instability region appears at smaller rotation rates.   

The results in Fig.~\ref{fig:tauA} make qualitative sense given our expectations for the low T/W instability, but we would obviously like to understand the detailed features better.  
There are two main aspects to this. First of all, the maximum $\omega_I$ is not centred 
in the middle of the co-rotation band, where $\varpi_{cor} = 0.5$. Secondly,  $\omega_I$ appears to reach its largest absolute value in the $A=0.4$ model.   

If the shear instability depends on an energy exchange between the 
bulk motion and the co-rotating mode, one would  expect  the instability timescale to depend  
on the energy and degree of differential rotation present at the co-rotation point. Considering the behaviour of the variables involved, 
we find that the  function;
\begin{equation}
\psi = - \beta \left.  \frac{\partial \Omega}{\partial \varpi} \right| _{\varpi_{cor}} \, ,  \label{eq:psi}
\end{equation}
where the derivative of the star's angular velocity is calculated at the co-rotation radius, has a maximum 
at the same  point as $\omega_I$ along the stellar sequence. Actually, we find that $\psi$ increases from $A=0.1$ to 0.5 and then decreases.  Hence, the function $\psi$   accounts reasonably well for the main features observed in our results, eg.
the faster instability growth of the model with $A\sim0.4$.  
Moreover, for each rotating model (with a given $A$)  the function $\psi$ can be used to determine
 the position of the shortest growth time inside the instability region. 
 
It is worth noting that $\psi$ can be re-written in terms 
 of the star's parameters and the mode's pattern speed, using equation~(\ref{eq:wc}), 
\begin{equation}
\psi = \frac{ 2 \beta }{A \Omega_c } \left(   \Omega_c - \sigma  \right) ^{\frac{1}{2}}  \sigma^{\frac{3}{2}} \, .  \label{eq:psiB}
\end{equation}
Our numerical data suggest that  $\omega_I$ is reasonably well  approximated  by the following empirical relation;  
\begin{equation}
 \chi = a \, \psi ^{10 A} \label{eq:chi} \, , 
\end{equation}
where $a$ is a constant, and $A$ is the usual differential rotation parameter.   
 For the various models considered in this work the function $\chi$ is shown in Fig.~\ref{fig:psiA}, while 
 the corresponding (dimensionless) values of $a$ are given in Table~\ref{tab:psiA}.
 The maximum and the behaviour at small co-rotation radii are well captured by $\chi$, there is a slight horizontal  offset 
 for $A=0.2$ and 0.6 but they only amount to 1-2 \% with respect to the $\omega_I$  peak.   
 For larger  $\varpi_{cor}$, the function $\chi$ generally assumes slightly larger values than  $\omega_I$, while  
 for  $A=0.1$  the opposite is the case. For the latter model, this behaviour could be due to the presence of other unstable modes, 
 but we did not manage to resolve this issue with our simulations.  

 Considering the explorative approach adopted in our investigation, it is interesting to note how well the function $\chi$ 
  captures the main properties of the instability growth time and how it identifies the model with the strongest instability along a sequence. 
  This provides interesting pointers towards future work. First of all, one should   
  investigate if our empirical expression remains useful also for other  models, eg. different polytropic indices 
  and rotational laws. If the suggestion proves robust one should also try to understand the underlying physics. This might shed useful light on the nature of this class of 
  instabilities. 

Before moving on, let us point out that most of the numerical results have been presented in dimensionless units, and the shortest growth time calculated in this work is 
 $\tau = 29.25 / \sqrt{G \rho_m }$ for a model with $A=0.4$ and axis ratio $R_p / R_{eq} = 0.4$. 
In physical units, this value is  $\tau = 5.13$~ms, for  a typical star with $M=1.4 M_{\odot}$ and 
EoS parameters $\gamma=2$ and 
$k=6.674 \cdot 10^4 \, \textrm{g}^{-1} \textrm{cm}^5 \textrm{s}^{-2} $.

\begin{table}
\begin{center}
\caption{\label{tab:psiA} 
The dimensionless values  
of the coefficient $a$, see equation (\ref{eq:chi}), for  stellar models with $A=0.1-0.6$.}
\begin{tabular}{c  c  c c c c c}
\hline
 $ A $  &  0.1 & 0.2 & 0.3 & 0.4 & 0.5 & 0.6  \\
 \hline  
 $ a $  &  0.072 & 0.74 & 2.32 & 5.26 & 12.4 & 32.9 \\
 \hline  
\end{tabular}
\end{center}
\end{table}

\begin{figure}
\begin{center}
\includegraphics[height=82mm]{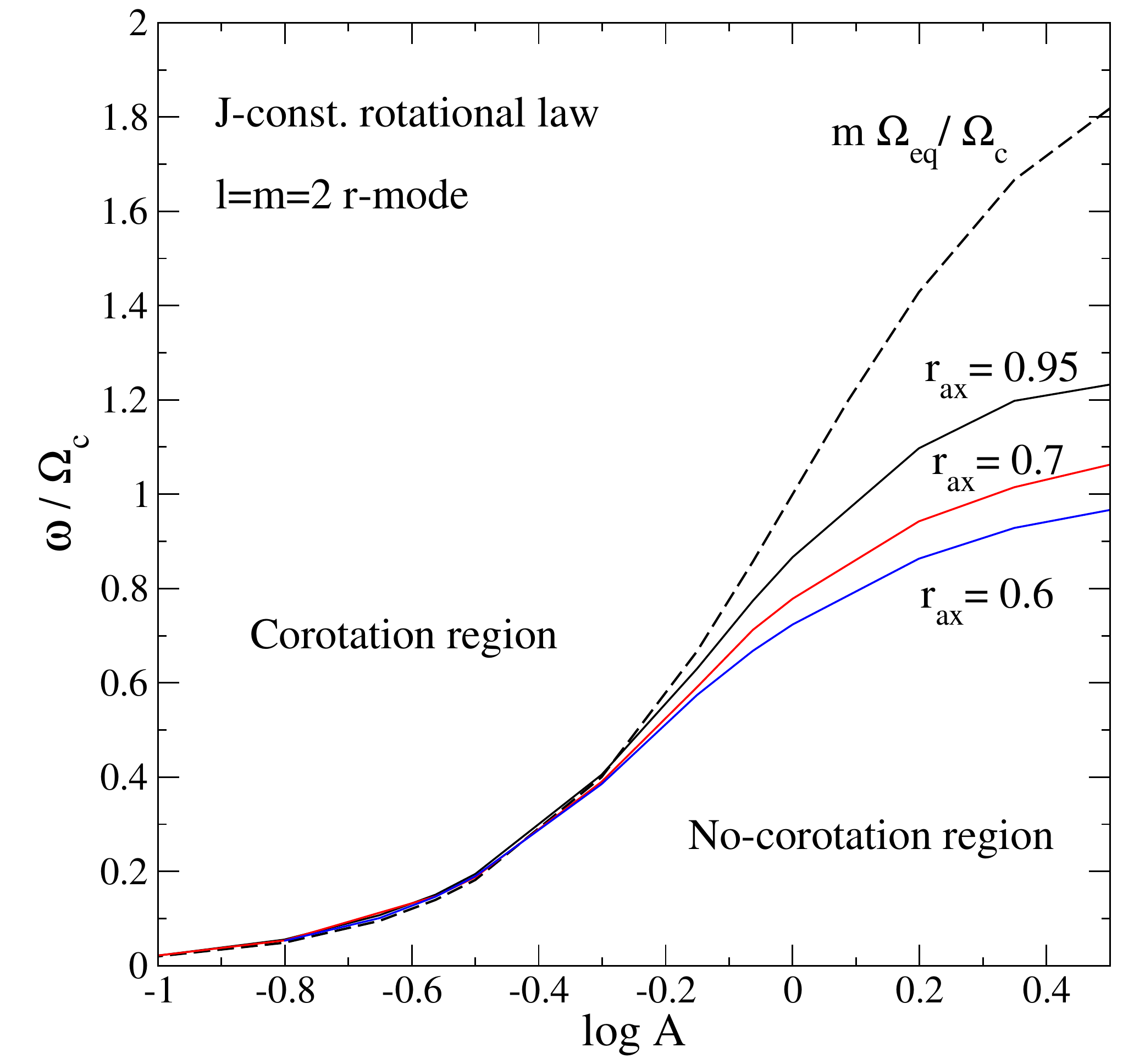} 
\caption{This figure shows the dependence of the $l=m=2$ r-mode on the differential rotation parameter $A$, for 
three stellar configurations with a fixed axis ratio. They are, respectively, $r_{ax}=0.95, 0.7$ and 0.6 (see legend). 
The dashed line  denotes the equatorial boundary of the co-rotation region. The horizontal axis displays the quantity $\log A$ while the 
vertical axis gives the ratio between the mode frequency and the central angular velocity of the star. This figure can be directly compared 
to Fig. 4 of \citet{2001PhRvD..64b4003K}, where the numerical method became unreliable as the mode approached the co-rotation region.
\label{fig:r-mode}}
\end{center}
\end{figure}

\section{A slight aside: The inertial r-modes } \label{sec:Engc}

The perturbative time-evolution code obviously allows us to consider a wider range of problems involving differentially rotating stars. As a pertinent example, let us make some comments on the inertial r-modes. These modes are rotationally restored by the Coriolis force and are of particular interest since they suffer the gravitational-wave instability already at low rotation rates in uniform stars. Much effort has gone into understanding the astrophysical relevance of this instability, eg. the role of various internal viscosity mechanisms, but the bulk of the work has considered stars in uniform rotation. If the r-mode instability acts very early on in a neutron star's life or is triggered in the hot remnant immediately after binary neutron star merger, then we would need to understand how differential rotation affects the mechanism. It is easy to see that the co-rotation point could be relevant for this discussion.  According to an inertial observer the r-mode is prograde for any stellar rotation rate with a frequency proportional to the bulk rotation rate (like all inertial modes). 
If such a mode enters the co-rotation region it should do so by crossing
 the low-frequency boundary  which  corresponds to the equatorial angular velocity $\Omega_{eq}$ (at least for ``reasonable'' rotation laws like the j-constant law).
 
 However, it is not clear from the  results in the literature if the r-mode actually enters the co-rotation region. So far, the r-mode frequency in differentially rotating stars  has only been determined in Newtonian gravity by~\citet{2001PhRvD..64b4003K}.  
For a $\gamma=2$ polytrope and the j-constant rotation law, their results show that the r-mode is not in co-rotation for stars with $A>0.6$. However, the mode clearly approaches the co-rotation band 
when $A\to 0.6$.  Unfortunately, the numerical code used by~\citet{2001PhRvD..64b4003K} was not accurate for higher degrees of differential rotation 
(presumably because the co-rotation point corresponds to a singularity in the frequency domain description of the problem). 
As our numerical framework is based on time evolving the perturbation equations, our analysis is not affect by this issue and we can  determine the r-mode frequency for lower 
values of $A$. This allows us to check if the r-mode goes in co-rotation for a high degree of differential rotation.  To carry out this exercise, 
we use the Cowling approximation which is known to provides accurate results for the r-mode frequency~\citep{2003MNRAS.343..175K}. 
Considering models with a fixed axis ratio, $r_{ax} = R_p/R_{eq}$, respectively,  $r_{ax} = 0.95, 0.7$ and 0.6, we determine the $l=m=2$ r-mode frequency 
for  $A=0.1-10$. The results are shown in Figure~\ref{fig:r-mode}. We find that, even though the r-mode approaches the co-rotation region as the degree of differential rotation is increased, 
 it never really enters the co-rotation region (more than marginally). In all cases we have considered, the r-mode stays   
very close to the equatorial boundary, i.e. $\omega \simeq m \Omega_{eq}$, as the rotation rate increases. We have found no evidence that the r-mode suffers a shearing instability analogous to the low T/W instability in the f-mode. It is not clear whether this is because no such instability exists or if it is simply the case that the growth time is too large to overcome the numerical dissipation in the examples we have studied (as one would expect for a mode located at the boundary of the co-rotation region).

\section{Concluding remarks\label{conclusions}} \label{sec:concl}

We have developed a framework for numerical time evolutions of linear perturbations  of differentially rotating, Newtonian, stars. Within this framework we have carried out a 
large parameter survey for
polytropic stars described by the standard j-constant rotation law, focussing on the fundamental f-mode and the low T/W instability. Our results for the instability growth time confirm the assertion of 
\cite{2005ApJ...618L..37W}  that the instability sets in at the point when the f-mode first enters the co-rotation region (when there is a point in the star where the pattern speed in the star matches the local rotation rate of the background fluid). We have used the integrands of the canonical energy and angular momentum to highlight the close relationship between the instability and the co-rotation point. By considering the relevant stellar parameters we have also obtained a relatively simple empirical relation which describes how the growth rate of the low T/W instability increases as the f-mode moves into the  co-rotation region, reaches a maximum and then tapers off as the mode exits the unstable region again. 

Our analysis has, in many ways, been of an exploratory nature. Our main aim was to map out the relevant parameter space and provide clear evidence for the link between the co-rotation boundary and the onset of instability. As they stand the results seem convincing in this respect. This is quite an important conclusion because  it would be sufficient to know that the f-mode of a given configuration lies inside the co-rotation region to know that the mode should be (at least in principle) unstable. Further work is needed to confirm this conclusion for other rotation laws and more realistic equations of state, but we have no reason to expect that the qualitative behaviour would change much in other cases. More detailed work is also needed in order to explain the form of the empirical relation we have found for the instability growth rate. Ideally, one would like to be able to derive this result from underlying principles, but this seems like an very difficult problem. One might be able to make progress using the canonical energy/angular momentum arguments from \citet{1975ApJ...200..204F, 1978ApJ...221..937F, 1978ApJ...222..281F}. However, this would likely require a frequency domain approach which would be hampered by the singularity associated with the co-rotation point. Unfortunately, the fact that the instability analysis only holds (strictly) for inviscid systems means that we have been unable to use our results to make progress in this direction (the numerical simulations involves the standard artificial viscosity to stabilize the evolutions). We have also not been able to extract the mode eigenfunctions with sufficient precision due to the exponential growth of the perturbation variables during the instability. It seems possible that future work could do better in this regard. We would certainly encourage efforts in this direction.

Finally, we used our computational framework to consider the inertial r-modes. In this case, our results extend previous work to higher degrees of differential rotation.  We have  demonstrated that the r-modes do not (at least for our stellar models) evolve into the co-rotation region. Rather, they remain near the boundary of this region for rapidly and differentially rotating stars. This is an interesting result in its own right, and a clear demonstration of the usefulness of our numerical framework. The result may also be relevant for relativistic studies of the r-mode where the frame-dragging provides an effect that is qualitatively similar to differential rotation. For non-barotropic stars, in particular, the relativistic r-mode problem is known to be complicated. Based on the present evidence it would be tempting to suggest that time-evolutions of the kind used in the present work may prove a powerful tool also in that setting.

\section*{Acknowledgements}
AP acknowledges support from the European Union Seventh
Framework Programme (FP7/2007-2013) under grant agreement
no. 267251 ``Astronomy Fellowships in Italy (AstroFIt)''. NA acknowledges support from STFC in the UK and generous hospitality from the Institute for Nuclear Theory at the University of Washington where some of this work was carried out.


\nocite*


\begin{thebibliography}{}

\bibitem[\protect\citeauthoryear{{Andersson}}{{Andersson}}{2003}]{2003CQGra..20R.105A}
{Andersson} N.,  2003, Class. and Quantum Grav., 20, 105

\bibitem[\protect\citeauthoryear{{Baiotti}, {Pietri}, {Manca} \&
  {Rezzolla}}{{Baiotti} et~al.}{2007}]{2007PhRvD..75d4023B}
{Baiotti} L.,  {Pietri} R.~D.,  {Manca} G.~M.,    {Rezzolla} L.,  2007, \prd,
  75, 044023

\bibitem[\protect\citeauthoryear{{Camarda}, {Anninos}, {Fragile} \&
  {Font}}{{Camarda} et~al.}{2009}]{2009ApJ...707.1610C}
{Camarda} K.~D.,  {Anninos} P.,  {Fragile} P.~C.,    {Font} J.~A.,  2009, \apj,
  707, 1610

\bibitem[\protect\citeauthoryear{{Centrella}, {New}, {Lowe} \&
  {Brown}}{{Centrella} et~al.}{2001}]{2001ApJ...550L.193C}
{Centrella} J.~M.,  {New} K.~C.~B.,  {Lowe} L.~L.,    {Brown} J.~D.,  2001,
  \apjl, 550, L193

\bibitem[\protect\citeauthoryear{{Cerd{\'a}-Dur{\'a}n}, {Quilis} \&
  {Font}}{{Cerd{\'a}-Dur{\'a}n} et~al.}{2007}]{2007CoPhC.177..288C}
{Cerd{\'a}-Dur{\'a}n} P.,  {Quilis} V.,    {Font} J.~A.,  2007, Computer
  Physics Communications, 177, 288

\bibitem[\protect\citeauthoryear{{Chandrasekhar}}{{Chandrasekhar}}{1969}]{1969efe..book.....C}
{Chandrasekhar} S.,  1969, {Ellipsoidal figures of equilibrium (New Haven, CT: Yale University Press)}


\bibitem[\protect\citeauthoryear{{Chandrasekhar}}{{Chandrasekhar}}{1970}]{1970PhRvL..24..611C}
{Chandrasekhar} S.,  1970, Phys. Rev. Lett., 24, 611

\bibitem[\protect\citeauthoryear{{Corvino}, {Rezzolla}, {Bernuzzi}, {De Pietri}
  \& {Giacomazzo}}{{Corvino} et~al.}{2010}]{2010CQGra..27k4104C}
{Corvino} G.,  {Rezzolla} L.,  {Bernuzzi} S.,  {De Pietri} R.,    {Giacomazzo}
  B.,  2010, Class. and Quantum Grav., 27, 114104

\bibitem[\protect\citeauthoryear{{Franci}, {De Pietri}, {Dionysopoulou} \&
  {Rezzolla}}{{Franci} et~al.}{2013}]{2013PhRvD..88j4028F}
{Franci} L.,  {De Pietri} R.,  {Dionysopoulou} K.,    {Rezzolla} L.,  2013,
  \prd, 88, 104028

\bibitem[\protect\citeauthoryear{{Friedman} \& {Schutz}}{{Friedman} \&
  {Schutz}}{1975}]{1975ApJ...200..204F}
{Friedman} J.~L.,  {Schutz} B.~F.,  1975, \apj, 200, 204

\bibitem[\protect\citeauthoryear{{Friedman} \& {Schutz}}{{Friedman} \&
  {Schutz}}{1978a}]{1978ApJ...221..937F}
{Friedman} J.~L.,  {Schutz} B.~F.,  1978a, \apj, 221, 937

\bibitem[\protect\citeauthoryear{{Friedman} \& {Schutz}}{{Friedman} \&
  {Schutz}}{1978b}]{1978ApJ...222..281F}
{Friedman} J.~L.,  {Schutz} B.~F.,  1978b, \apj, 222, 281

\bibitem[\protect\citeauthoryear{{Fu} \& {Lai}}{{Fu} \&
  {Lai}}{2011}]{2011MNRAS.413.2207F}
{Fu} W.,  {Lai} D.,  2011, \mnras, 413, 2207

\bibitem[\protect\citeauthoryear{{Hachisu}}{{Hachisu}}{1986}]{1986ApJS...61..479H}
{Hachisu} I.,  1986, \apj SS, 61, 479

\bibitem[\protect\citeauthoryear{{Jones}, {Andersson} \& {Stergioulas}}{{Jones}
  et~al.}{2002}]{2002MNRAS.334..933J}
{Jones} D.~I.,  {Andersson} N.,    {Stergioulas} N.,  2002, \mnras, 334, 933

\bibitem[\protect\citeauthoryear{{Karino}}{{Karino}}{2003}]{2003MNRAS.343..175K}
{Karino} S.,  2003, \mnras, 343, 175

\bibitem[\protect\citeauthoryear{Karino, Yoshida \& Eriguchi}{Karino
  et~al.}{2001}]{PhysRevD.64.024003}
Karino S.,  Yoshida S.,    Eriguchi Y.,  2001, Phys. Rev. D, 64, 024003

\bibitem[\protect\citeauthoryear{{Karino}, {Yoshida} \& {Eriguchi}}{{Karino}
  et~al.}{2001}]{2001PhRvD..64b4003K}
{Karino} S.,  {Yoshida} S.,    {Eriguchi} Y.,  2001, \prd, 64, 024003

\bibitem[\protect\citeauthoryear{{Kuroda} \& {Umeda}}{{Kuroda} \&
  {Umeda}}{2010}]{2010ApJS..191..439K}
{Kuroda} T.,  {Umeda} H.,  2010, \apjs, 191, 439

\bibitem[\protect\citeauthoryear{{Muhlberger}, {Nouri}, {Duez}, {Foucart},
  {Kidder}, {Ott}, {Scheel}, {Szil{\'a}gyi} \& {Teukolsky}}{{Muhlberger}
  et~al.}{2014}]{2014arXiv1405.2144M}
{Muhlberger} C.~D.,  {Nouri} F.~H.,  {Duez} M.~D.,  {Foucart} F.,  {Kidder}
  L.~E.,  {Ott} C.~D.,  {Scheel} M.~A.,  {Szil{\'a}gyi} B.,    {Teukolsky}
  S.~A.,  2014, ArXiv:1405.2144

\bibitem[\protect\citeauthoryear{{Ott}, {Dimmelmeier}, {Marek}, {Janka},
  {Hawke}, {Zink} \& {Schnetter}}{{Ott} et~al.}{2007}]{2007PhRvL..98z1101O}
{Ott} C.~D.,  {Dimmelmeier} H.,  {Marek} A.,  {Janka} H.-T.,  {Hawke} I.,
  {Zink} B.,    {Schnetter} E.,  2007, Phys. Rev. Lett., 98, 261101

\bibitem[\protect\citeauthoryear{{Ott}, {Ou}, {Tohline} \& {Burrows}}{{Ott}
  et~al.}{2005}]{2005ApJ...625L.119O}
{Ott} C.~D.,  {Ou} S.,  {Tohline} J.~E.,    {Burrows} A.,  2005, \apjl, 625,
  L119

\bibitem[\protect\citeauthoryear{{Ou} \& {Tohline}}{{Ou} \&
  {Tohline}}{2006}]{2006ApJ...651.1068O}
{Ou} S.,  {Tohline} J.~E.,  2006, \apj, 651, 1068

\bibitem[\protect\citeauthoryear{{Papaloizou} \& {Pringle}}{{Papaloizou} \&
  {Pringle}}{1980}]{1980MNRAS.190...43P}
{Papaloizou} J.~C.,  {Pringle} J.~E.,  1980, \mnras, 190, 43

\bibitem[\protect\citeauthoryear{{Papaloizou} \& {Pringle}}{{Papaloizou} \&
  {Pringle}}{1984}]{1984MNRAS.208..721P}
{Papaloizou} J.~C.~B.,  {Pringle} J.~E.,  1984, \mnras, 208, 721


\bibitem[\protect\citeauthoryear{{Passamonti}, {Haskell}, {Andersson}, {Jones}
  \& {Hawke}}{{Passamonti} et~al.}{2009a}]{2009MNRAS.394..730P}
{Passamonti} A.,  {Haskell} B.,  {Andersson} N.,  {Jones} D.~I.,    {Hawke} I.,
   2009a, \mnras, 394, 730

\bibitem[\protect\citeauthoryear{{Passamonti}, {Haskell} \&
  {Andersson}}{{Passamonti} et~al.}{2009b}]{2009MNRAS.396..951P}
{Passamonti} A.,  {Haskell} B.,    {Andersson} N.,  2009b, \mnras, 396, 951


\bibitem[\protect\citeauthoryear{{Saijo}, {Baumgarte} \& {Shapiro}}{{Saijo}
  et~al.}{2003}]{2003ApJ...595..352S}
{Saijo} M.,  {Baumgarte} T.~W.,    {Shapiro} S.~L.,  2003, \apj, 595, 352

\bibitem[\protect\citeauthoryear{{Saijo} \& {Yoshida}}{{Saijo} \&
  {Yoshida}}{2006}]{2006MNRAS.368.1429S}
{Saijo} M.,  {Yoshida} S.,  2006, \mnras, 368, 1429

\bibitem[\protect\citeauthoryear{{Scheidegger}, {Fischer}, {Whitehouse} \&
  {Liebend{\"o}rfer}}{{Scheidegger} et~al.}{2008}]{2008A&A...490..231S}
{Scheidegger} S.,  {Fischer} T.,  {Whitehouse} S.~C.,    {Liebend{\"o}rfer} M.,
   2008, \aap, 490, 231

\bibitem[\protect\citeauthoryear{{Schutz}}{{Schutz}}{1980a}]{1980MNRAS.190...21S}
{Schutz} B.~F.,  1980a, \mnras, 190, 21

\bibitem[\protect\citeauthoryear{{Schutz}}{{Schutz}}{1980b}]{1980MNRAS.190....7S}
{Schutz} B.~F.,  1980b, \mnras, 190, 7

\bibitem[\protect\citeauthoryear{{Shibata}, {Baumgarte} \& {Shapiro}}{{Shibata}
  et~al.}{2000}]{2000ApJ...542..453S}
{Shibata} M.,  {Baumgarte} T.~W.,    {Shapiro} S.~L.,  2000, \apj, 542, 453

\bibitem[\protect\citeauthoryear{{Shibata}, {Karino} \& {Eriguchi}}{{Shibata}
  et~al.}{2002}]{2002MNRAS.334L..27S}
{Shibata} M.,  {Karino} S.,    {Eriguchi} Y.,  2002, \mnras, 334, L27

\bibitem[\protect\citeauthoryear{{Shibata}, {Karino} \& {Eriguchi}}{{Shibata}
  et~al.}{2003}]{2003MNRAS.343..619S}
{Shibata} M.,  {Karino} S.,    {Eriguchi} Y.,  2003, \mnras, 343, 619

\bibitem[\protect\citeauthoryear{{Watts}, {Andersson} \& {Jones}}{{Watts}
  et~al.}{2005}]{2005ApJ...618L..37W}
{Watts} A.~L.,  {Andersson} N.,    {Jones} D.~I.,  2005, \apjl, 618, L37

\end{thebibliography}

\label{lastpage}
\end{document}